\documentclass[smallextended]{svjour3}       % onecolumn (second format)

\smartqed  % flush right qed marks, e.g. at end of proof
\usepackage{graphicx}

\usepackage{epsfig,amsfonts,amssymb}
\usepackage{hyperref}
\usepackage{cite}
\input epsf.sty
\usepackage{cite}

\def\ZZZ{{\hbox{ Z\kern-1.6mm Z}}}
\def\RRR{{\hbox{ R\kern-2.4mm R}}}
\def\CCC{{\hbox{ C\kern-2.0mm C}}}
\def\zzz{{\hbox{z\kern-1mm z}}}

\newcommand{\qeq}{{\hbox{=\kern-2.3mm ? \kern.5mm }}}
\renewcommand{\qeq}{=}

\newcommand{\OO}{{\cal O}}

\newcommand{\NN}{{\cal N}}

\newcommand{\be}{\begin{equation}}
\newcommand{\ee}{\end{equation}}
\newcommand{\ben}{\begin{eqnarray}\displaystyle}
\newcommand{\een}{\end{eqnarray}}

\newcommand{\refb}[1]{(\ref{#1})}

\def\one{{\hbox{ 1\kern-.8mm l}}}
\def\zero{{\hbox{ 0\kern-1.5mm 0}}}

\newcommand{\bea}[1]{\begin{eqnarray}\label{#1} }
\newcommand{\eea}{\end{eqnarray}}

\newcommand{\eqref}{\refb}

\def\figadssmall{
\def\JPicScale{0.3}
\ifx\JPicScale\undefined\def\JPicScale{1}\fi
\unitlength \JPicScale mm
\begin{picture}(132.48,302.48)(0,0)
\linethickness{0.3mm}
\multiput(30,50)(0.49,0.1){1}{\line(1,0){0.49}}
\multiput(30.49,50.1)(0.49,0.1){1}{\line(1,0){0.49}}
\multiput(30.98,50.2)(0.49,0.1){1}{\line(1,0){0.49}}
\multiput(31.47,50.3)(0.49,0.1){1}{\line(1,0){0.49}}
\multiput(31.95,50.41)(0.49,0.11){1}{\line(1,0){0.49}}
\multiput(32.44,50.51)(0.49,0.11){1}{\line(1,0){0.49}}
\multiput(32.93,50.62)(0.49,0.11){1}{\line(1,0){0.49}}
\multiput(33.42,50.73)(0.49,0.11){1}{\line(1,0){0.49}}
\multiput(33.9,50.84)(0.49,0.11){1}{\line(1,0){0.49}}
\multiput(34.39,50.96)(0.49,0.12){1}{\line(1,0){0.49}}
\multiput(34.87,51.07)(0.48,0.12){1}{\line(1,0){0.48}}
\multiput(35.36,51.19)(0.48,0.12){1}{\line(1,0){0.48}}
\multiput(35.84,51.31)(0.48,0.12){1}{\line(1,0){0.48}}
\multiput(36.33,51.43)(0.48,0.12){1}{\line(1,0){0.48}}
\multiput(36.81,51.56)(0.48,0.13){1}{\line(1,0){0.48}}
\multiput(37.29,51.68)(0.48,0.13){1}{\line(1,0){0.48}}
\multiput(37.78,51.81)(0.48,0.13){1}{\line(1,0){0.48}}
\multiput(38.26,51.94)(0.48,0.13){1}{\line(1,0){0.48}}
\multiput(38.74,52.07)(0.48,0.13){1}{\line(1,0){0.48}}
\multiput(39.22,52.2)(0.48,0.13){1}{\line(1,0){0.48}}
\multiput(39.7,52.34)(0.48,0.14){1}{\line(1,0){0.48}}
\multiput(40.18,52.48)(0.48,0.14){1}{\line(1,0){0.48}}
\multiput(40.66,52.61)(0.48,0.14){1}{\line(1,0){0.48}}
\multiput(41.14,52.75)(0.48,0.14){1}{\line(1,0){0.48}}
\multiput(41.62,52.9)(0.48,0.14){1}{\line(1,0){0.48}}
\multiput(42.1,53.04)(0.48,0.15){1}{\line(1,0){0.48}}
\multiput(42.57,53.19)(0.48,0.15){1}{\line(1,0){0.48}}
\multiput(43.05,53.34)(0.48,0.15){1}{\line(1,0){0.48}}
\multiput(43.53,53.49)(0.48,0.15){1}{\line(1,0){0.48}}
\multiput(44,53.64)(0.47,0.15){1}{\line(1,0){0.47}}
\multiput(44.48,53.79)(0.47,0.16){1}{\line(1,0){0.47}}
\multiput(44.95,53.95)(0.47,0.16){1}{\line(1,0){0.47}}
\multiput(45.42,54.1)(0.47,0.16){1}{\line(1,0){0.47}}
\multiput(45.9,54.26)(0.47,0.16){1}{\line(1,0){0.47}}
\multiput(46.37,54.42)(0.47,0.16){1}{\line(1,0){0.47}}
\multiput(46.84,54.59)(0.47,0.16){1}{\line(1,0){0.47}}
\multiput(47.31,54.75)(0.47,0.17){1}{\line(1,0){0.47}}
\multiput(47.78,54.92)(0.47,0.17){1}{\line(1,0){0.47}}
\multiput(48.25,55.09)(0.47,0.17){1}{\line(1,0){0.47}}
\multiput(48.72,55.26)(0.47,0.17){1}{\line(1,0){0.47}}
\multiput(49.19,55.43)(0.47,0.17){1}{\line(1,0){0.47}}
\multiput(49.66,55.6)(0.47,0.18){1}{\line(1,0){0.47}}
\multiput(50.13,55.78)(0.47,0.18){1}{\line(1,0){0.47}}
\multiput(50.59,55.96)(0.47,0.18){1}{\line(1,0){0.47}}
\multiput(51.06,56.14)(0.23,0.09){2}{\line(1,0){0.23}}
\multiput(51.52,56.32)(0.23,0.09){2}{\line(1,0){0.23}}
\multiput(51.99,56.5)(0.23,0.09){2}{\line(1,0){0.23}}
\multiput(52.45,56.69)(0.23,0.09){2}{\line(1,0){0.23}}
\multiput(52.91,56.87)(0.23,0.09){2}{\line(1,0){0.23}}
\multiput(53.38,57.06)(0.23,0.1){2}{\line(1,0){0.23}}
\multiput(53.84,57.25)(0.23,0.1){2}{\line(1,0){0.23}}
\multiput(54.3,57.44)(0.23,0.1){2}{\line(1,0){0.23}}
\multiput(54.76,57.64)(0.23,0.1){2}{\line(1,0){0.23}}
\multiput(55.22,57.83)(0.23,0.1){2}{\line(1,0){0.23}}
\multiput(55.68,58.03)(0.23,0.1){2}{\line(1,0){0.23}}
\multiput(56.13,58.23)(0.23,0.1){2}{\line(1,0){0.23}}
\multiput(56.59,58.43)(0.23,0.1){2}{\line(1,0){0.23}}
\multiput(57.05,58.63)(0.23,0.1){2}{\line(1,0){0.23}}
\multiput(57.5,58.84)(0.23,0.1){2}{\line(1,0){0.23}}
\multiput(57.96,59.04)(0.23,0.1){2}{\line(1,0){0.23}}
\multiput(58.41,59.25)(0.23,0.11){2}{\line(1,0){0.23}}
\multiput(58.86,59.46)(0.23,0.11){2}{\line(1,0){0.23}}
\multiput(59.31,59.67)(0.23,0.11){2}{\line(1,0){0.23}}
\multiput(59.77,59.89)(0.23,0.11){2}{\line(1,0){0.23}}
\multiput(60.22,60.1)(0.22,0.11){2}{\line(1,0){0.22}}
\multiput(60.67,60.32)(0.22,0.11){2}{\line(1,0){0.22}}
\multiput(61.11,60.54)(0.22,0.11){2}{\line(1,0){0.22}}
\multiput(61.56,60.76)(0.22,0.11){2}{\line(1,0){0.22}}
\multiput(62.01,60.98)(0.22,0.11){2}{\line(1,0){0.22}}
\multiput(62.45,61.21)(0.22,0.11){2}{\line(1,0){0.22}}
\multiput(62.9,61.43)(0.22,0.11){2}{\line(1,0){0.22}}
\multiput(63.34,61.66)(0.22,0.11){2}{\line(1,0){0.22}}
\multiput(63.79,61.89)(0.22,0.12){2}{\line(1,0){0.22}}
\multiput(64.23,62.12)(0.22,0.12){2}{\line(1,0){0.22}}
\multiput(64.67,62.35)(0.22,0.12){2}{\line(1,0){0.22}}
\multiput(65.11,62.59)(0.22,0.12){2}{\line(1,0){0.22}}
\multiput(65.55,62.82)(0.22,0.12){2}{\line(1,0){0.22}}
\multiput(65.99,63.06)(0.22,0.12){2}{\line(1,0){0.22}}
\multiput(66.43,63.3)(0.22,0.12){2}{\line(1,0){0.22}}
\multiput(66.87,63.54)(0.22,0.12){2}{\line(1,0){0.22}}
\multiput(67.3,63.79)(0.22,0.12){2}{\line(1,0){0.22}}
\multiput(67.74,64.03)(0.22,0.12){2}{\line(1,0){0.22}}
\multiput(68.17,64.28)(0.22,0.12){2}{\line(1,0){0.22}}
\multiput(68.6,64.53)(0.22,0.12){2}{\line(1,0){0.22}}
\multiput(69.04,64.78)(0.22,0.13){2}{\line(1,0){0.22}}
\multiput(69.47,65.03)(0.22,0.13){2}{\line(1,0){0.22}}
\multiput(69.9,65.28)(0.21,0.13){2}{\line(1,0){0.21}}
\multiput(70.33,65.54)(0.21,0.13){2}{\line(1,0){0.21}}
\multiput(70.75,65.79)(0.21,0.13){2}{\line(1,0){0.21}}
\multiput(71.18,66.05)(0.21,0.13){2}{\line(1,0){0.21}}
\multiput(71.61,66.31)(0.21,0.13){2}{\line(1,0){0.21}}
\multiput(72.03,66.57)(0.21,0.13){2}{\line(1,0){0.21}}
\multiput(72.46,66.83)(0.21,0.13){2}{\line(1,0){0.21}}
\multiput(72.88,67.1)(0.21,0.13){2}{\line(1,0){0.21}}
\multiput(73.3,67.37)(0.21,0.13){2}{\line(1,0){0.21}}
\multiput(73.72,67.63)(0.21,0.13){2}{\line(1,0){0.21}}
\multiput(74.14,67.9)(0.21,0.14){2}{\line(1,0){0.21}}
\multiput(74.56,68.18)(0.21,0.14){2}{\line(1,0){0.21}}
\multiput(74.98,68.45)(0.21,0.14){2}{\line(1,0){0.21}}
\multiput(75.39,68.72)(0.21,0.14){2}{\line(1,0){0.21}}
\multiput(75.81,69)(0.21,0.14){2}{\line(1,0){0.21}}
\multiput(76.22,69.28)(0.21,0.14){2}{\line(1,0){0.21}}
\multiput(76.64,69.56)(0.21,0.14){2}{\line(1,0){0.21}}
\multiput(77.05,69.84)(0.21,0.14){2}{\line(1,0){0.21}}
\multiput(77.46,70.12)(0.21,0.14){2}{\line(1,0){0.21}}
\multiput(77.87,70.41)(0.2,0.14){2}{\line(1,0){0.2}}
\multiput(78.28,70.69)(0.2,0.14){2}{\line(1,0){0.2}}
\multiput(78.69,70.98)(0.2,0.14){2}{\line(1,0){0.2}}
\multiput(79.1,71.27)(0.2,0.15){2}{\line(1,0){0.2}}
\multiput(79.5,71.56)(0.2,0.15){2}{\line(1,0){0.2}}
\multiput(79.91,71.85)(0.2,0.15){2}{\line(1,0){0.2}}
\multiput(80.31,72.15)(0.2,0.15){2}{\line(1,0){0.2}}
\multiput(80.71,72.44)(0.2,0.15){2}{\line(1,0){0.2}}
\multiput(81.11,72.74)(0.2,0.15){2}{\line(1,0){0.2}}
\multiput(81.51,73.04)(0.13,0.1){3}{\line(1,0){0.13}}
\multiput(81.91,73.34)(0.13,0.1){3}{\line(1,0){0.13}}
\multiput(82.31,73.64)(0.13,0.1){3}{\line(1,0){0.13}}
\multiput(82.7,73.95)(0.13,0.1){3}{\line(1,0){0.13}}
\multiput(83.1,74.25)(0.13,0.1){3}{\line(1,0){0.13}}
\multiput(83.49,74.56)(0.13,0.1){3}{\line(1,0){0.13}}
\multiput(83.89,74.86)(0.13,0.1){3}{\line(1,0){0.13}}
\multiput(84.28,75.17)(0.13,0.1){3}{\line(1,0){0.13}}
\multiput(84.67,75.49)(0.13,0.1){3}{\line(1,0){0.13}}
\multiput(85.06,75.8)(0.13,0.1){3}{\line(1,0){0.13}}
\multiput(85.44,76.11)(0.13,0.11){3}{\line(1,0){0.13}}
\multiput(85.83,76.43)(0.13,0.11){3}{\line(1,0){0.13}}
\multiput(86.22,76.75)(0.13,0.11){3}{\line(1,0){0.13}}
\multiput(86.6,77.06)(0.13,0.11){3}{\line(1,0){0.13}}
\multiput(86.98,77.38)(0.13,0.11){3}{\line(1,0){0.13}}
\multiput(87.36,77.71)(0.13,0.11){3}{\line(1,0){0.13}}
\multiput(87.74,78.03)(0.13,0.11){3}{\line(1,0){0.13}}
\multiput(88.12,78.35)(0.13,0.11){3}{\line(1,0){0.13}}
\multiput(88.5,78.68)(0.13,0.11){3}{\line(1,0){0.13}}
\multiput(88.88,79.01)(0.13,0.11){3}{\line(1,0){0.13}}
\multiput(89.25,79.34)(0.12,0.11){3}{\line(1,0){0.12}}
\multiput(89.63,79.67)(0.12,0.11){3}{\line(1,0){0.12}}
\linethickness{0.3mm}
\multiput(89.63,10.33)(0.12,-0.11){3}{\line(1,0){0.12}}
\multiput(89.25,10.66)(0.12,-0.11){3}{\line(1,0){0.12}}
\multiput(88.88,10.99)(0.13,-0.11){3}{\line(1,0){0.13}}
\multiput(88.5,11.32)(0.13,-0.11){3}{\line(1,0){0.13}}
\multiput(88.12,11.65)(0.13,-0.11){3}{\line(1,0){0.13}}
\multiput(87.74,11.97)(0.13,-0.11){3}{\line(1,0){0.13}}
\multiput(87.36,12.29)(0.13,-0.11){3}{\line(1,0){0.13}}
\multiput(86.98,12.62)(0.13,-0.11){3}{\line(1,0){0.13}}
\multiput(86.6,12.94)(0.13,-0.11){3}{\line(1,0){0.13}}
\multiput(86.22,13.25)(0.13,-0.11){3}{\line(1,0){0.13}}
\multiput(85.83,13.57)(0.13,-0.11){3}{\line(1,0){0.13}}
\multiput(85.44,13.89)(0.13,-0.11){3}{\line(1,0){0.13}}
\multiput(85.06,14.2)(0.13,-0.1){3}{\line(1,0){0.13}}
\multiput(84.67,14.51)(0.13,-0.1){3}{\line(1,0){0.13}}
\multiput(84.28,14.83)(0.13,-0.1){3}{\line(1,0){0.13}}
\multiput(83.89,15.14)(0.13,-0.1){3}{\line(1,0){0.13}}
\multiput(83.49,15.44)(0.13,-0.1){3}{\line(1,0){0.13}}
\multiput(83.1,15.75)(0.13,-0.1){3}{\line(1,0){0.13}}
\multiput(82.7,16.05)(0.13,-0.1){3}{\line(1,0){0.13}}
\multiput(82.31,16.36)(0.13,-0.1){3}{\line(1,0){0.13}}
\multiput(81.91,16.66)(0.13,-0.1){3}{\line(1,0){0.13}}
\multiput(81.51,16.96)(0.13,-0.1){3}{\line(1,0){0.13}}
\multiput(81.11,17.26)(0.2,-0.15){2}{\line(1,0){0.2}}
\multiput(80.71,17.56)(0.2,-0.15){2}{\line(1,0){0.2}}
\multiput(80.31,17.85)(0.2,-0.15){2}{\line(1,0){0.2}}
\multiput(79.91,18.15)(0.2,-0.15){2}{\line(1,0){0.2}}
\multiput(79.5,18.44)(0.2,-0.15){2}{\line(1,0){0.2}}
\multiput(79.1,18.73)(0.2,-0.15){2}{\line(1,0){0.2}}
\multiput(78.69,19.02)(0.2,-0.14){2}{\line(1,0){0.2}}
\multiput(78.28,19.31)(0.2,-0.14){2}{\line(1,0){0.2}}
\multiput(77.87,19.59)(0.2,-0.14){2}{\line(1,0){0.2}}
\multiput(77.46,19.88)(0.21,-0.14){2}{\line(1,0){0.21}}
\multiput(77.05,20.16)(0.21,-0.14){2}{\line(1,0){0.21}}
\multiput(76.64,20.44)(0.21,-0.14){2}{\line(1,0){0.21}}
\multiput(76.22,20.72)(0.21,-0.14){2}{\line(1,0){0.21}}
\multiput(75.81,21)(0.21,-0.14){2}{\line(1,0){0.21}}
\multiput(75.39,21.28)(0.21,-0.14){2}{\line(1,0){0.21}}
\multiput(74.98,21.55)(0.21,-0.14){2}{\line(1,0){0.21}}
\multiput(74.56,21.82)(0.21,-0.14){2}{\line(1,0){0.21}}
\multiput(74.14,22.1)(0.21,-0.14){2}{\line(1,0){0.21}}
\multiput(73.72,22.37)(0.21,-0.13){2}{\line(1,0){0.21}}
\multiput(73.3,22.63)(0.21,-0.13){2}{\line(1,0){0.21}}
\multiput(72.88,22.9)(0.21,-0.13){2}{\line(1,0){0.21}}
\multiput(72.46,23.17)(0.21,-0.13){2}{\line(1,0){0.21}}
\multiput(72.03,23.43)(0.21,-0.13){2}{\line(1,0){0.21}}
\multiput(71.61,23.69)(0.21,-0.13){2}{\line(1,0){0.21}}
\multiput(71.18,23.95)(0.21,-0.13){2}{\line(1,0){0.21}}
\multiput(70.75,24.21)(0.21,-0.13){2}{\line(1,0){0.21}}
\multiput(70.33,24.46)(0.21,-0.13){2}{\line(1,0){0.21}}
\multiput(69.9,24.72)(0.21,-0.13){2}{\line(1,0){0.21}}
\multiput(69.47,24.97)(0.22,-0.13){2}{\line(1,0){0.22}}
\multiput(69.04,25.22)(0.22,-0.13){2}{\line(1,0){0.22}}
\multiput(68.6,25.47)(0.22,-0.12){2}{\line(1,0){0.22}}
\multiput(68.17,25.72)(0.22,-0.12){2}{\line(1,0){0.22}}
\multiput(67.74,25.97)(0.22,-0.12){2}{\line(1,0){0.22}}
\multiput(67.3,26.21)(0.22,-0.12){2}{\line(1,0){0.22}}
\multiput(66.87,26.46)(0.22,-0.12){2}{\line(1,0){0.22}}
\multiput(66.43,26.7)(0.22,-0.12){2}{\line(1,0){0.22}}
\multiput(65.99,26.94)(0.22,-0.12){2}{\line(1,0){0.22}}
\multiput(65.55,27.18)(0.22,-0.12){2}{\line(1,0){0.22}}
\multiput(65.11,27.41)(0.22,-0.12){2}{\line(1,0){0.22}}
\multiput(64.67,27.65)(0.22,-0.12){2}{\line(1,0){0.22}}
\multiput(64.23,27.88)(0.22,-0.12){2}{\line(1,0){0.22}}
\multiput(63.79,28.11)(0.22,-0.12){2}{\line(1,0){0.22}}
\multiput(63.34,28.34)(0.22,-0.11){2}{\line(1,0){0.22}}
\multiput(62.9,28.57)(0.22,-0.11){2}{\line(1,0){0.22}}
\multiput(62.45,28.79)(0.22,-0.11){2}{\line(1,0){0.22}}
\multiput(62.01,29.02)(0.22,-0.11){2}{\line(1,0){0.22}}
\multiput(61.56,29.24)(0.22,-0.11){2}{\line(1,0){0.22}}
\multiput(61.11,29.46)(0.22,-0.11){2}{\line(1,0){0.22}}
\multiput(60.67,29.68)(0.22,-0.11){2}{\line(1,0){0.22}}
\multiput(60.22,29.9)(0.22,-0.11){2}{\line(1,0){0.22}}
\multiput(59.77,30.11)(0.23,-0.11){2}{\line(1,0){0.23}}
\multiput(59.31,30.33)(0.23,-0.11){2}{\line(1,0){0.23}}
\multiput(58.86,30.54)(0.23,-0.11){2}{\line(1,0){0.23}}
\multiput(58.41,30.75)(0.23,-0.11){2}{\line(1,0){0.23}}
\multiput(57.96,30.96)(0.23,-0.1){2}{\line(1,0){0.23}}
\multiput(57.5,31.16)(0.23,-0.1){2}{\line(1,0){0.23}}
\multiput(57.05,31.37)(0.23,-0.1){2}{\line(1,0){0.23}}
\multiput(56.59,31.57)(0.23,-0.1){2}{\line(1,0){0.23}}
\multiput(56.13,31.77)(0.23,-0.1){2}{\line(1,0){0.23}}
\multiput(55.68,31.97)(0.23,-0.1){2}{\line(1,0){0.23}}
\multiput(55.22,32.17)(0.23,-0.1){2}{\line(1,0){0.23}}
\multiput(54.76,32.36)(0.23,-0.1){2}{\line(1,0){0.23}}
\multiput(54.3,32.56)(0.23,-0.1){2}{\line(1,0){0.23}}
\multiput(53.84,32.75)(0.23,-0.1){2}{\line(1,0){0.23}}
\multiput(53.38,32.94)(0.23,-0.1){2}{\line(1,0){0.23}}
\multiput(52.91,33.13)(0.23,-0.09){2}{\line(1,0){0.23}}
\multiput(52.45,33.31)(0.23,-0.09){2}{\line(1,0){0.23}}
\multiput(51.99,33.5)(0.23,-0.09){2}{\line(1,0){0.23}}
\multiput(51.52,33.68)(0.23,-0.09){2}{\line(1,0){0.23}}
\multiput(51.06,33.86)(0.23,-0.09){2}{\line(1,0){0.23}}
\multiput(50.59,34.04)(0.47,-0.18){1}{\line(1,0){0.47}}
\multiput(50.13,34.22)(0.47,-0.18){1}{\line(1,0){0.47}}
\multiput(49.66,34.4)(0.47,-0.18){1}{\line(1,0){0.47}}
\multiput(49.19,34.57)(0.47,-0.17){1}{\line(1,0){0.47}}
\multiput(48.72,34.74)(0.47,-0.17){1}{\line(1,0){0.47}}
\multiput(48.25,34.91)(0.47,-0.17){1}{\line(1,0){0.47}}
\multiput(47.78,35.08)(0.47,-0.17){1}{\line(1,0){0.47}}
\multiput(47.31,35.25)(0.47,-0.17){1}{\line(1,0){0.47}}
\multiput(46.84,35.41)(0.47,-0.16){1}{\line(1,0){0.47}}
\multiput(46.37,35.58)(0.47,-0.16){1}{\line(1,0){0.47}}
\multiput(45.9,35.74)(0.47,-0.16){1}{\line(1,0){0.47}}
\multiput(45.42,35.9)(0.47,-0.16){1}{\line(1,0){0.47}}
\multiput(44.95,36.05)(0.47,-0.16){1}{\line(1,0){0.47}}
\multiput(44.48,36.21)(0.47,-0.16){1}{\line(1,0){0.47}}
\multiput(44,36.36)(0.47,-0.15){1}{\line(1,0){0.47}}
\multiput(43.53,36.51)(0.48,-0.15){1}{\line(1,0){0.48}}
\multiput(43.05,36.66)(0.48,-0.15){1}{\line(1,0){0.48}}
\multiput(42.57,36.81)(0.48,-0.15){1}{\line(1,0){0.48}}
\multiput(42.1,36.96)(0.48,-0.15){1}{\line(1,0){0.48}}
\multiput(41.62,37.1)(0.48,-0.14){1}{\line(1,0){0.48}}
\multiput(41.14,37.25)(0.48,-0.14){1}{\line(1,0){0.48}}
\multiput(40.66,37.39)(0.48,-0.14){1}{\line(1,0){0.48}}
\multiput(40.18,37.52)(0.48,-0.14){1}{\line(1,0){0.48}}
\multiput(39.7,37.66)(0.48,-0.14){1}{\line(1,0){0.48}}
\multiput(39.22,37.8)(0.48,-0.13){1}{\line(1,0){0.48}}
\multiput(38.74,37.93)(0.48,-0.13){1}{\line(1,0){0.48}}
\multiput(38.26,38.06)(0.48,-0.13){1}{\line(1,0){0.48}}
\multiput(37.78,38.19)(0.48,-0.13){1}{\line(1,0){0.48}}
\multiput(37.29,38.32)(0.48,-0.13){1}{\line(1,0){0.48}}
\multiput(36.81,38.44)(0.48,-0.13){1}{\line(1,0){0.48}}
\multiput(36.33,38.57)(0.48,-0.12){1}{\line(1,0){0.48}}
\multiput(35.84,38.69)(0.48,-0.12){1}{\line(1,0){0.48}}
\multiput(35.36,38.81)(0.48,-0.12){1}{\line(1,0){0.48}}
\multiput(34.87,38.93)(0.48,-0.12){1}{\line(1,0){0.48}}
\multiput(34.39,39.04)(0.49,-0.12){1}{\line(1,0){0.49}}
\multiput(33.9,39.16)(0.49,-0.11){1}{\line(1,0){0.49}}
\multiput(33.42,39.27)(0.49,-0.11){1}{\line(1,0){0.49}}
\multiput(32.93,39.38)(0.49,-0.11){1}{\line(1,0){0.49}}
\multiput(32.44,39.49)(0.49,-0.11){1}{\line(1,0){0.49}}
\multiput(31.95,39.59)(0.49,-0.11){1}{\line(1,0){0.49}}
\multiput(31.47,39.7)(0.49,-0.1){1}{\line(1,0){0.49}}
\multiput(30.98,39.8)(0.49,-0.1){1}{\line(1,0){0.49}}
\multiput(30.49,39.9)(0.49,-0.1){1}{\line(1,0){0.49}}
\multiput(30,40)(0.49,-0.1){1}{\line(1,0){0.49}}
\linethickness{0.3mm}
\put(27.5,44.75){\line(0,1){0.49}}
\multiput(27.5,44.75)(0.02,-0.49){1}{\line(0,-1){0.49}}
\multiput(27.53,44.27)(0.05,-0.48){1}{\line(0,-1){0.48}}
\multiput(27.57,43.79)(0.07,-0.47){1}{\line(0,-1){0.47}}
\multiput(27.65,43.32)(0.09,-0.45){1}{\line(0,-1){0.45}}
\multiput(27.74,42.86)(0.12,-0.43){1}{\line(0,-1){0.43}}
\multiput(27.86,42.43)(0.14,-0.41){1}{\line(0,-1){0.41}}
\multiput(27.99,42.02)(0.16,-0.38){1}{\line(0,-1){0.38}}
\multiput(28.15,41.64)(0.17,-0.35){1}{\line(0,-1){0.35}}
\multiput(28.32,41.3)(0.09,-0.16){2}{\line(0,-1){0.16}}
\multiput(28.51,40.98)(0.1,-0.14){2}{\line(0,-1){0.14}}
\multiput(28.71,40.71)(0.11,-0.12){2}{\line(0,-1){0.12}}
\multiput(28.93,40.48)(0.11,-0.09){2}{\line(1,0){0.11}}
\multiput(29.16,40.29)(0.23,-0.14){1}{\line(1,0){0.23}}
\multiput(29.39,40.15)(0.24,-0.1){1}{\line(1,0){0.24}}
\multiput(29.63,40.05)(0.24,-0.05){1}{\line(1,0){0.24}}
\put(29.88,40.01){\line(1,0){0.25}}
\multiput(30.12,40.01)(0.24,0.05){1}{\line(1,0){0.24}}
\multiput(30.37,40.05)(0.24,0.1){1}{\line(1,0){0.24}}
\multiput(30.61,40.15)(0.23,0.14){1}{\line(1,0){0.23}}
\multiput(30.84,40.29)(0.11,0.09){2}{\line(1,0){0.11}}
\multiput(31.07,40.48)(0.11,0.12){2}{\line(0,1){0.12}}
\multiput(31.29,40.71)(0.1,0.14){2}{\line(0,1){0.14}}
\multiput(31.49,40.98)(0.09,0.16){2}{\line(0,1){0.16}}
\multiput(31.68,41.3)(0.17,0.35){1}{\line(0,1){0.35}}
\multiput(31.85,41.64)(0.16,0.38){1}{\line(0,1){0.38}}
\multiput(32.01,42.02)(0.14,0.41){1}{\line(0,1){0.41}}
\multiput(32.14,42.43)(0.12,0.43){1}{\line(0,1){0.43}}
\multiput(32.26,42.86)(0.09,0.45){1}{\line(0,1){0.45}}
\multiput(32.35,43.32)(0.07,0.47){1}{\line(0,1){0.47}}
\multiput(32.43,43.79)(0.05,0.48){1}{\line(0,1){0.48}}
\multiput(32.47,44.27)(0.02,0.49){1}{\line(0,1){0.49}}
\put(32.5,44.75){\line(0,1){0.49}}
\multiput(32.47,45.73)(0.02,-0.49){1}{\line(0,-1){0.49}}
\multiput(32.43,46.21)(0.05,-0.48){1}{\line(0,-1){0.48}}
\multiput(32.35,46.68)(0.07,-0.47){1}{\line(0,-1){0.47}}
\multiput(32.26,47.14)(0.09,-0.45){1}{\line(0,-1){0.45}}
\multiput(32.14,47.57)(0.12,-0.43){1}{\line(0,-1){0.43}}
\multiput(32.01,47.98)(0.14,-0.41){1}{\line(0,-1){0.41}}
\multiput(31.85,48.36)(0.16,-0.38){1}{\line(0,-1){0.38}}
\multiput(31.68,48.7)(0.17,-0.35){1}{\line(0,-1){0.35}}
\multiput(31.49,49.02)(0.09,-0.16){2}{\line(0,-1){0.16}}
\multiput(31.29,49.29)(0.1,-0.14){2}{\line(0,-1){0.14}}
\multiput(31.07,49.52)(0.11,-0.12){2}{\line(0,-1){0.12}}
\multiput(30.84,49.71)(0.11,-0.09){2}{\line(1,0){0.11}}
\multiput(30.61,49.85)(0.23,-0.14){1}{\line(1,0){0.23}}
\multiput(30.37,49.95)(0.24,-0.1){1}{\line(1,0){0.24}}
\multiput(30.12,49.99)(0.24,-0.05){1}{\line(1,0){0.24}}
\put(29.88,49.99){\line(1,0){0.25}}
\multiput(29.63,49.95)(0.24,0.05){1}{\line(1,0){0.24}}
\multiput(29.39,49.85)(0.24,0.1){1}{\line(1,0){0.24}}
\multiput(29.16,49.71)(0.23,0.14){1}{\line(1,0){0.23}}
\multiput(28.93,49.52)(0.11,0.09){2}{\line(1,0){0.11}}
\multiput(28.71,49.29)(0.11,0.12){2}{\line(0,1){0.12}}
\multiput(28.51,49.02)(0.1,0.14){2}{\line(0,1){0.14}}
\multiput(28.32,48.7)(0.09,0.16){2}{\line(0,1){0.16}}
\multiput(28.15,48.36)(0.17,0.35){1}{\line(0,1){0.35}}
\multiput(27.99,47.98)(0.16,0.38){1}{\line(0,1){0.38}}
\multiput(27.86,47.57)(0.14,0.41){1}{\line(0,1){0.41}}
\multiput(27.74,47.14)(0.12,0.43){1}{\line(0,1){0.43}}
\multiput(27.65,46.68)(0.09,0.45){1}{\line(0,1){0.45}}
\multiput(27.57,46.21)(0.07,0.47){1}{\line(0,1){0.47}}
\multiput(27.53,45.73)(0.05,0.48){1}{\line(0,1){0.48}}
\multiput(27.5,45.25)(0.02,0.49){1}{\line(0,1){0.49}}
\linethickness{0.3mm}
\put(87.5,44.75){\line(0,1){0.5}}
\multiput(87.5,44.75)(0,-0.5){1}{\line(0,-1){0.5}}
\multiput(87.5,44.25)(0,-0.5){1}{\line(0,-1){0.5}}
\multiput(87.5,43.75)(0,-0.5){1}{\line(0,-1){0.5}}
\multiput(87.5,43.25)(0,-0.5){1}{\line(0,-1){0.5}}
\multiput(87.51,42.75)(0,-0.5){1}{\line(0,-1){0.5}}
\multiput(87.51,42.25)(0,-0.5){1}{\line(0,-1){0.5}}
\multiput(87.51,41.76)(0,-0.5){1}{\line(0,-1){0.5}}
\multiput(87.51,41.26)(0,-0.5){1}{\line(0,-1){0.5}}
\multiput(87.52,40.76)(0,-0.5){1}{\line(0,-1){0.5}}
\multiput(87.52,40.27)(0.01,-0.49){1}{\line(0,-1){0.49}}
\multiput(87.53,39.77)(0.01,-0.49){1}{\line(0,-1){0.49}}
\multiput(87.53,39.28)(0.01,-0.49){1}{\line(0,-1){0.49}}
\multiput(87.54,38.79)(0.01,-0.49){1}{\line(0,-1){0.49}}
\multiput(87.55,38.29)(0.01,-0.49){1}{\line(0,-1){0.49}}
\multiput(87.55,37.8)(0.01,-0.49){1}{\line(0,-1){0.49}}
\multiput(87.56,37.32)(0.01,-0.49){1}{\line(0,-1){0.49}}
\multiput(87.57,36.83)(0.01,-0.49){1}{\line(0,-1){0.49}}
\multiput(87.58,36.34)(0.01,-0.48){1}{\line(0,-1){0.48}}
\multiput(87.59,35.86)(0.01,-0.48){1}{\line(0,-1){0.48}}
\multiput(87.6,35.38)(0.01,-0.48){1}{\line(0,-1){0.48}}
\multiput(87.61,34.9)(0.01,-0.48){1}{\line(0,-1){0.48}}
\multiput(87.62,34.42)(0.01,-0.48){1}{\line(0,-1){0.48}}
\multiput(87.63,33.95)(0.01,-0.47){1}{\line(0,-1){0.47}}
\multiput(87.64,33.47)(0.01,-0.47){1}{\line(0,-1){0.47}}
\multiput(87.65,33)(0.01,-0.47){1}{\line(0,-1){0.47}}
\multiput(87.66,32.53)(0.01,-0.47){1}{\line(0,-1){0.47}}
\multiput(87.68,32.07)(0.01,-0.46){1}{\line(0,-1){0.46}}
\multiput(87.69,31.61)(0.01,-0.46){1}{\line(0,-1){0.46}}
\multiput(87.7,31.15)(0.01,-0.46){1}{\line(0,-1){0.46}}
\multiput(87.72,30.69)(0.01,-0.45){1}{\line(0,-1){0.45}}
\multiput(87.73,30.23)(0.02,-0.45){1}{\line(0,-1){0.45}}
\multiput(87.75,29.78)(0.02,-0.45){1}{\line(0,-1){0.45}}
\multiput(87.76,29.33)(0.02,-0.45){1}{\line(0,-1){0.45}}
\multiput(87.78,28.89)(0.02,-0.44){1}{\line(0,-1){0.44}}
\multiput(87.8,28.45)(0.02,-0.44){1}{\line(0,-1){0.44}}
\multiput(87.81,28.01)(0.02,-0.44){1}{\line(0,-1){0.44}}
\multiput(87.83,27.57)(0.02,-0.43){1}{\line(0,-1){0.43}}
\multiput(87.85,27.14)(0.02,-0.43){1}{\line(0,-1){0.43}}
\multiput(87.87,26.71)(0.02,-0.42){1}{\line(0,-1){0.42}}
\multiput(87.89,26.29)(0.02,-0.42){1}{\line(0,-1){0.42}}
\multiput(87.91,25.87)(0.02,-0.42){1}{\line(0,-1){0.42}}
\multiput(87.93,25.45)(0.02,-0.41){1}{\line(0,-1){0.41}}
\multiput(87.95,25.04)(0.02,-0.41){1}{\line(0,-1){0.41}}
\multiput(87.97,24.63)(0.02,-0.4){1}{\line(0,-1){0.4}}
\multiput(87.99,24.23)(0.02,-0.4){1}{\line(0,-1){0.4}}
\multiput(88.01,23.83)(0.02,-0.4){1}{\line(0,-1){0.4}}
\multiput(88.03,23.43)(0.02,-0.39){1}{\line(0,-1){0.39}}
\multiput(88.05,23.04)(0.02,-0.39){1}{\line(0,-1){0.39}}
\multiput(88.08,22.65)(0.02,-0.38){1}{\line(0,-1){0.38}}
\multiput(88.1,22.27)(0.02,-0.38){1}{\line(0,-1){0.38}}
\multiput(88.12,21.89)(0.02,-0.37){1}{\line(0,-1){0.37}}
\multiput(88.15,21.52)(0.02,-0.37){1}{\line(0,-1){0.37}}
\multiput(88.17,21.15)(0.02,-0.36){1}{\line(0,-1){0.36}}
\multiput(88.19,20.79)(0.02,-0.36){1}{\line(0,-1){0.36}}
\multiput(88.22,20.43)(0.03,-0.35){1}{\line(0,-1){0.35}}
\multiput(88.24,20.08)(0.03,-0.35){1}{\line(0,-1){0.35}}
\multiput(88.27,19.73)(0.03,-0.34){1}{\line(0,-1){0.34}}
\multiput(88.3,19.38)(0.03,-0.34){1}{\line(0,-1){0.34}}
\multiput(88.32,19.05)(0.03,-0.33){1}{\line(0,-1){0.33}}
\multiput(88.35,18.71)(0.03,-0.33){1}{\line(0,-1){0.33}}
\multiput(88.38,18.39)(0.03,-0.32){1}{\line(0,-1){0.32}}
\multiput(88.4,18.06)(0.03,-0.32){1}{\line(0,-1){0.32}}
\multiput(88.43,17.75)(0.03,-0.31){1}{\line(0,-1){0.31}}
\multiput(88.46,17.44)(0.03,-0.31){1}{\line(0,-1){0.31}}
\multiput(88.49,17.13)(0.03,-0.3){1}{\line(0,-1){0.3}}
\multiput(88.52,16.83)(0.03,-0.29){1}{\line(0,-1){0.29}}
\multiput(88.55,16.54)(0.03,-0.29){1}{\line(0,-1){0.29}}
\multiput(88.57,16.25)(0.03,-0.28){1}{\line(0,-1){0.28}}
\multiput(88.6,15.97)(0.03,-0.28){1}{\line(0,-1){0.28}}
\multiput(88.63,15.69)(0.03,-0.27){1}{\line(0,-1){0.27}}
\multiput(88.66,15.42)(0.03,-0.26){1}{\line(0,-1){0.26}}
\multiput(88.69,15.16)(0.03,-0.26){1}{\line(0,-1){0.26}}
\multiput(88.72,14.9)(0.03,-0.25){1}{\line(0,-1){0.25}}
\multiput(88.76,14.65)(0.03,-0.25){1}{\line(0,-1){0.25}}
\multiput(88.79,14.4)(0.03,-0.24){1}{\line(0,-1){0.24}}
\multiput(88.82,14.16)(0.03,-0.23){1}{\line(0,-1){0.23}}
\multiput(88.85,13.93)(0.03,-0.23){1}{\line(0,-1){0.23}}
\multiput(88.88,13.7)(0.03,-0.22){1}{\line(0,-1){0.22}}
\multiput(88.91,13.48)(0.03,-0.21){1}{\line(0,-1){0.21}}
\multiput(88.95,13.27)(0.03,-0.21){1}{\line(0,-1){0.21}}
\multiput(88.98,13.06)(0.03,-0.2){1}{\line(0,-1){0.2}}
\multiput(89.01,12.86)(0.03,-0.19){1}{\line(0,-1){0.19}}
\multiput(89.04,12.66)(0.03,-0.19){1}{\line(0,-1){0.19}}
\multiput(89.08,12.48)(0.03,-0.18){1}{\line(0,-1){0.18}}
\multiput(89.11,12.29)(0.03,-0.17){1}{\line(0,-1){0.17}}
\multiput(89.14,12.12)(0.03,-0.17){1}{\line(0,-1){0.17}}
\multiput(89.18,11.95)(0.03,-0.16){1}{\line(0,-1){0.16}}
\multiput(89.21,11.79)(0.03,-0.15){1}{\line(0,-1){0.15}}
\multiput(89.24,11.64)(0.03,-0.15){1}{\line(0,-1){0.15}}
\multiput(89.28,11.49)(0.03,-0.14){1}{\line(0,-1){0.14}}
\multiput(89.31,11.35)(0.03,-0.13){1}{\line(0,-1){0.13}}
\multiput(89.35,11.21)(0.03,-0.13){1}{\line(0,-1){0.13}}
\multiput(89.38,11.09)(0.03,-0.12){1}{\line(0,-1){0.12}}
\multiput(89.42,10.97)(0.03,-0.11){1}{\line(0,-1){0.11}}
\multiput(89.45,10.85)(0.03,-0.11){1}{\line(0,-1){0.11}}
\multiput(89.49,10.75)(0.03,-0.1){1}{\line(0,-1){0.1}}
\multiput(89.52,10.65)(0.04,-0.09){1}{\line(0,-1){0.09}}
\multiput(89.56,10.56)(0.04,-0.09){1}{\line(0,-1){0.09}}
\multiput(89.59,10.47)(0.04,-0.08){1}{\line(0,-1){0.08}}
\multiput(89.63,10.39)(0.04,-0.07){1}{\line(0,-1){0.07}}
\multiput(89.66,10.32)(0.04,-0.06){1}{\line(0,-1){0.06}}
\multiput(89.7,10.26)(0.04,-0.06){1}{\line(0,-1){0.06}}
\multiput(89.73,10.2)(0.04,-0.05){1}{\line(0,-1){0.05}}
\multiput(89.77,10.15)(0.04,-0.04){1}{\line(0,-1){0.04}}
\multiput(89.8,10.11)(0.04,-0.04){1}{\line(0,-1){0.04}}
\multiput(89.84,10.07)(0.04,-0.03){1}{\line(1,0){0.04}}
\multiput(89.88,10.04)(0.04,-0.02){1}{\line(1,0){0.04}}
\multiput(89.91,10.02)(0.04,-0.01){1}{\line(1,0){0.04}}
\multiput(89.95,10.01)(0.04,-0.01){1}{\line(1,0){0.04}}
\put(89.98,10){\line(1,0){0.04}}
\multiput(90.02,10)(0.04,0.01){1}{\line(1,0){0.04}}
\multiput(90.05,10.01)(0.04,0.01){1}{\line(1,0){0.04}}
\multiput(90.09,10.02)(0.04,0.02){1}{\line(1,0){0.04}}
\multiput(90.12,10.04)(0.04,0.03){1}{\line(1,0){0.04}}
\multiput(90.16,10.07)(0.04,0.04){1}{\line(0,1){0.04}}
\multiput(90.2,10.11)(0.04,0.04){1}{\line(0,1){0.04}}
\multiput(90.23,10.15)(0.04,0.05){1}{\line(0,1){0.05}}
\multiput(90.27,10.2)(0.04,0.06){1}{\line(0,1){0.06}}
\multiput(90.3,10.26)(0.04,0.06){1}{\line(0,1){0.06}}
\multiput(90.34,10.32)(0.04,0.07){1}{\line(0,1){0.07}}
\multiput(90.37,10.39)(0.04,0.08){1}{\line(0,1){0.08}}
\multiput(90.41,10.47)(0.04,0.09){1}{\line(0,1){0.09}}
\multiput(90.44,10.56)(0.04,0.09){1}{\line(0,1){0.09}}
\multiput(90.48,10.65)(0.03,0.1){1}{\line(0,1){0.1}}
\multiput(90.51,10.75)(0.03,0.11){1}{\line(0,1){0.11}}
\multiput(90.55,10.85)(0.03,0.11){1}{\line(0,1){0.11}}
\multiput(90.58,10.97)(0.03,0.12){1}{\line(0,1){0.12}}
\multiput(90.62,11.09)(0.03,0.13){1}{\line(0,1){0.13}}
\multiput(90.65,11.21)(0.03,0.13){1}{\line(0,1){0.13}}
\multiput(90.69,11.35)(0.03,0.14){1}{\line(0,1){0.14}}
\multiput(90.72,11.49)(0.03,0.15){1}{\line(0,1){0.15}}
\multiput(90.76,11.64)(0.03,0.15){1}{\line(0,1){0.15}}
\multiput(90.79,11.79)(0.03,0.16){1}{\line(0,1){0.16}}
\multiput(90.82,11.95)(0.03,0.17){1}{\line(0,1){0.17}}
\multiput(90.86,12.12)(0.03,0.17){1}{\line(0,1){0.17}}
\multiput(90.89,12.29)(0.03,0.18){1}{\line(0,1){0.18}}
\multiput(90.92,12.48)(0.03,0.19){1}{\line(0,1){0.19}}
\multiput(90.96,12.66)(0.03,0.19){1}{\line(0,1){0.19}}
\multiput(90.99,12.86)(0.03,0.2){1}{\line(0,1){0.2}}
\multiput(91.02,13.06)(0.03,0.21){1}{\line(0,1){0.21}}
\multiput(91.05,13.27)(0.03,0.21){1}{\line(0,1){0.21}}
\multiput(91.09,13.48)(0.03,0.22){1}{\line(0,1){0.22}}
\multiput(91.12,13.7)(0.03,0.23){1}{\line(0,1){0.23}}
\multiput(91.15,13.93)(0.03,0.23){1}{\line(0,1){0.23}}
\multiput(91.18,14.16)(0.03,0.24){1}{\line(0,1){0.24}}
\multiput(91.21,14.4)(0.03,0.25){1}{\line(0,1){0.25}}
\multiput(91.24,14.65)(0.03,0.25){1}{\line(0,1){0.25}}
\multiput(91.28,14.9)(0.03,0.26){1}{\line(0,1){0.26}}
\multiput(91.31,15.16)(0.03,0.26){1}{\line(0,1){0.26}}
\multiput(91.34,15.42)(0.03,0.27){1}{\line(0,1){0.27}}
\multiput(91.37,15.69)(0.03,0.28){1}{\line(0,1){0.28}}
\multiput(91.4,15.97)(0.03,0.28){1}{\line(0,1){0.28}}
\multiput(91.43,16.25)(0.03,0.29){1}{\line(0,1){0.29}}
\multiput(91.45,16.54)(0.03,0.29){1}{\line(0,1){0.29}}
\multiput(91.48,16.83)(0.03,0.3){1}{\line(0,1){0.3}}
\multiput(91.51,17.13)(0.03,0.31){1}{\line(0,1){0.31}}
\multiput(91.54,17.44)(0.03,0.31){1}{\line(0,1){0.31}}
\multiput(91.57,17.75)(0.03,0.32){1}{\line(0,1){0.32}}
\multiput(91.6,18.06)(0.03,0.32){1}{\line(0,1){0.32}}
\multiput(91.62,18.39)(0.03,0.33){1}{\line(0,1){0.33}}
\multiput(91.65,18.71)(0.03,0.33){1}{\line(0,1){0.33}}
\multiput(91.68,19.05)(0.03,0.34){1}{\line(0,1){0.34}}
\multiput(91.7,19.38)(0.03,0.34){1}{\line(0,1){0.34}}
\multiput(91.73,19.73)(0.03,0.35){1}{\line(0,1){0.35}}
\multiput(91.76,20.08)(0.03,0.35){1}{\line(0,1){0.35}}
\multiput(91.78,20.43)(0.02,0.36){1}{\line(0,1){0.36}}
\multiput(91.81,20.79)(0.02,0.36){1}{\line(0,1){0.36}}
\multiput(91.83,21.15)(0.02,0.37){1}{\line(0,1){0.37}}
\multiput(91.85,21.52)(0.02,0.37){1}{\line(0,1){0.37}}
\multiput(91.88,21.89)(0.02,0.38){1}{\line(0,1){0.38}}
\multiput(91.9,22.27)(0.02,0.38){1}{\line(0,1){0.38}}
\multiput(91.92,22.65)(0.02,0.39){1}{\line(0,1){0.39}}
\multiput(91.95,23.04)(0.02,0.39){1}{\line(0,1){0.39}}
\multiput(91.97,23.43)(0.02,0.4){1}{\line(0,1){0.4}}
\multiput(91.99,23.83)(0.02,0.4){1}{\line(0,1){0.4}}
\multiput(92.01,24.23)(0.02,0.4){1}{\line(0,1){0.4}}
\multiput(92.03,24.63)(0.02,0.41){1}{\line(0,1){0.41}}
\multiput(92.05,25.04)(0.02,0.41){1}{\line(0,1){0.41}}
\multiput(92.07,25.45)(0.02,0.42){1}{\line(0,1){0.42}}
\multiput(92.09,25.87)(0.02,0.42){1}{\line(0,1){0.42}}
\multiput(92.11,26.29)(0.02,0.42){1}{\line(0,1){0.42}}
\multiput(92.13,26.71)(0.02,0.43){1}{\line(0,1){0.43}}
\multiput(92.15,27.14)(0.02,0.43){1}{\line(0,1){0.43}}
\multiput(92.17,27.57)(0.02,0.44){1}{\line(0,1){0.44}}
\multiput(92.19,28.01)(0.02,0.44){1}{\line(0,1){0.44}}
\multiput(92.2,28.45)(0.02,0.44){1}{\line(0,1){0.44}}
\multiput(92.22,28.89)(0.02,0.45){1}{\line(0,1){0.45}}
\multiput(92.24,29.33)(0.02,0.45){1}{\line(0,1){0.45}}
\multiput(92.25,29.78)(0.02,0.45){1}{\line(0,1){0.45}}
\multiput(92.27,30.23)(0.01,0.45){1}{\line(0,1){0.45}}
\multiput(92.28,30.69)(0.01,0.46){1}{\line(0,1){0.46}}
\multiput(92.3,31.15)(0.01,0.46){1}{\line(0,1){0.46}}
\multiput(92.31,31.61)(0.01,0.46){1}{\line(0,1){0.46}}
\multiput(92.32,32.07)(0.01,0.47){1}{\line(0,1){0.47}}
\multiput(92.34,32.53)(0.01,0.47){1}{\line(0,1){0.47}}
\multiput(92.35,33)(0.01,0.47){1}{\line(0,1){0.47}}
\multiput(92.36,33.47)(0.01,0.47){1}{\line(0,1){0.47}}
\multiput(92.37,33.95)(0.01,0.48){1}{\line(0,1){0.48}}
\multiput(92.38,34.42)(0.01,0.48){1}{\line(0,1){0.48}}
\multiput(92.39,34.9)(0.01,0.48){1}{\line(0,1){0.48}}
\multiput(92.4,35.38)(0.01,0.48){1}{\line(0,1){0.48}}
\multiput(92.41,35.86)(0.01,0.48){1}{\line(0,1){0.48}}
\multiput(92.42,36.34)(0.01,0.49){1}{\line(0,1){0.49}}
\multiput(92.43,36.83)(0.01,0.49){1}{\line(0,1){0.49}}
\multiput(92.44,37.32)(0.01,0.49){1}{\line(0,1){0.49}}
\multiput(92.45,37.8)(0.01,0.49){1}{\line(0,1){0.49}}
\multiput(92.45,38.29)(0.01,0.49){1}{\line(0,1){0.49}}
\multiput(92.46,38.79)(0.01,0.49){1}{\line(0,1){0.49}}
\multiput(92.47,39.28)(0.01,0.49){1}{\line(0,1){0.49}}
\multiput(92.47,39.77)(0.01,0.49){1}{\line(0,1){0.49}}
\multiput(92.48,40.27)(0,0.5){1}{\line(0,1){0.5}}
\multiput(92.48,40.76)(0,0.5){1}{\line(0,1){0.5}}
\multiput(92.49,41.26)(0,0.5){1}{\line(0,1){0.5}}
\multiput(92.49,41.76)(0,0.5){1}{\line(0,1){0.5}}
\multiput(92.49,42.25)(0,0.5){1}{\line(0,1){0.5}}
\multiput(92.49,42.75)(0,0.5){1}{\line(0,1){0.5}}
\multiput(92.5,43.25)(0,0.5){1}{\line(0,1){0.5}}
\multiput(92.5,43.75)(0,0.5){1}{\line(0,1){0.5}}
\multiput(92.5,44.25)(0,0.5){1}{\line(0,1){0.5}}
\put(92.5,44.75){\line(0,1){0.5}}
\multiput(92.5,45.75)(0,-0.5){1}{\line(0,-1){0.5}}
\multiput(92.5,46.25)(0,-0.5){1}{\line(0,-1){0.5}}
\multiput(92.5,46.75)(0,-0.5){1}{\line(0,-1){0.5}}
\multiput(92.49,47.25)(0,-0.5){1}{\line(0,-1){0.5}}
\multiput(92.49,47.75)(0,-0.5){1}{\line(0,-1){0.5}}
\multiput(92.49,48.24)(0,-0.5){1}{\line(0,-1){0.5}}
\multiput(92.49,48.74)(0,-0.5){1}{\line(0,-1){0.5}}
\multiput(92.48,49.24)(0,-0.5){1}{\line(0,-1){0.5}}
\multiput(92.48,49.73)(0,-0.5){1}{\line(0,-1){0.5}}
\multiput(92.47,50.23)(0.01,-0.49){1}{\line(0,-1){0.49}}
\multiput(92.47,50.72)(0.01,-0.49){1}{\line(0,-1){0.49}}
\multiput(92.46,51.21)(0.01,-0.49){1}{\line(0,-1){0.49}}
\multiput(92.45,51.71)(0.01,-0.49){1}{\line(0,-1){0.49}}
\multiput(92.45,52.2)(0.01,-0.49){1}{\line(0,-1){0.49}}
\multiput(92.44,52.68)(0.01,-0.49){1}{\line(0,-1){0.49}}
\multiput(92.43,53.17)(0.01,-0.49){1}{\line(0,-1){0.49}}
\multiput(92.42,53.66)(0.01,-0.49){1}{\line(0,-1){0.49}}
\multiput(92.41,54.14)(0.01,-0.48){1}{\line(0,-1){0.48}}
\multiput(92.4,54.62)(0.01,-0.48){1}{\line(0,-1){0.48}}
\multiput(92.39,55.1)(0.01,-0.48){1}{\line(0,-1){0.48}}
\multiput(92.38,55.58)(0.01,-0.48){1}{\line(0,-1){0.48}}
\multiput(92.37,56.05)(0.01,-0.48){1}{\line(0,-1){0.48}}
\multiput(92.36,56.53)(0.01,-0.47){1}{\line(0,-1){0.47}}
\multiput(92.35,57)(0.01,-0.47){1}{\line(0,-1){0.47}}
\multiput(92.34,57.47)(0.01,-0.47){1}{\line(0,-1){0.47}}
\multiput(92.32,57.93)(0.01,-0.47){1}{\line(0,-1){0.47}}
\multiput(92.31,58.39)(0.01,-0.46){1}{\line(0,-1){0.46}}
\multiput(92.3,58.85)(0.01,-0.46){1}{\line(0,-1){0.46}}
\multiput(92.28,59.31)(0.01,-0.46){1}{\line(0,-1){0.46}}
\multiput(92.27,59.77)(0.01,-0.45){1}{\line(0,-1){0.45}}
\multiput(92.25,60.22)(0.02,-0.45){1}{\line(0,-1){0.45}}
\multiput(92.24,60.67)(0.02,-0.45){1}{\line(0,-1){0.45}}
\multiput(92.22,61.11)(0.02,-0.45){1}{\line(0,-1){0.45}}
\multiput(92.2,61.55)(0.02,-0.44){1}{\line(0,-1){0.44}}
\multiput(92.19,61.99)(0.02,-0.44){1}{\line(0,-1){0.44}}
\multiput(92.17,62.43)(0.02,-0.44){1}{\line(0,-1){0.44}}
\multiput(92.15,62.86)(0.02,-0.43){1}{\line(0,-1){0.43}}
\multiput(92.13,63.29)(0.02,-0.43){1}{\line(0,-1){0.43}}
\multiput(92.11,63.71)(0.02,-0.42){1}{\line(0,-1){0.42}}
\multiput(92.09,64.13)(0.02,-0.42){1}{\line(0,-1){0.42}}
\multiput(92.07,64.55)(0.02,-0.42){1}{\line(0,-1){0.42}}
\multiput(92.05,64.96)(0.02,-0.41){1}{\line(0,-1){0.41}}
\multiput(92.03,65.37)(0.02,-0.41){1}{\line(0,-1){0.41}}
\multiput(92.01,65.77)(0.02,-0.4){1}{\line(0,-1){0.4}}
\multiput(91.99,66.17)(0.02,-0.4){1}{\line(0,-1){0.4}}
\multiput(91.97,66.57)(0.02,-0.4){1}{\line(0,-1){0.4}}
\multiput(91.95,66.96)(0.02,-0.39){1}{\line(0,-1){0.39}}
\multiput(91.92,67.35)(0.02,-0.39){1}{\line(0,-1){0.39}}
\multiput(91.9,67.73)(0.02,-0.38){1}{\line(0,-1){0.38}}
\multiput(91.88,68.11)(0.02,-0.38){1}{\line(0,-1){0.38}}
\multiput(91.85,68.48)(0.02,-0.37){1}{\line(0,-1){0.37}}
\multiput(91.83,68.85)(0.02,-0.37){1}{\line(0,-1){0.37}}
\multiput(91.81,69.21)(0.02,-0.36){1}{\line(0,-1){0.36}}
\multiput(91.78,69.57)(0.02,-0.36){1}{\line(0,-1){0.36}}
\multiput(91.76,69.92)(0.03,-0.35){1}{\line(0,-1){0.35}}
\multiput(91.73,70.27)(0.03,-0.35){1}{\line(0,-1){0.35}}
\multiput(91.7,70.62)(0.03,-0.34){1}{\line(0,-1){0.34}}
\multiput(91.68,70.95)(0.03,-0.34){1}{\line(0,-1){0.34}}
\multiput(91.65,71.29)(0.03,-0.33){1}{\line(0,-1){0.33}}
\multiput(91.62,71.61)(0.03,-0.33){1}{\line(0,-1){0.33}}
\multiput(91.6,71.94)(0.03,-0.32){1}{\line(0,-1){0.32}}
\multiput(91.57,72.25)(0.03,-0.32){1}{\line(0,-1){0.32}}
\multiput(91.54,72.56)(0.03,-0.31){1}{\line(0,-1){0.31}}
\multiput(91.51,72.87)(0.03,-0.31){1}{\line(0,-1){0.31}}
\multiput(91.48,73.17)(0.03,-0.3){1}{\line(0,-1){0.3}}
\multiput(91.45,73.46)(0.03,-0.29){1}{\line(0,-1){0.29}}
\multiput(91.43,73.75)(0.03,-0.29){1}{\line(0,-1){0.29}}
\multiput(91.4,74.03)(0.03,-0.28){1}{\line(0,-1){0.28}}
\multiput(91.37,74.31)(0.03,-0.28){1}{\line(0,-1){0.28}}
\multiput(91.34,74.58)(0.03,-0.27){1}{\line(0,-1){0.27}}
\multiput(91.31,74.84)(0.03,-0.26){1}{\line(0,-1){0.26}}
\multiput(91.28,75.1)(0.03,-0.26){1}{\line(0,-1){0.26}}
\multiput(91.24,75.35)(0.03,-0.25){1}{\line(0,-1){0.25}}
\multiput(91.21,75.6)(0.03,-0.25){1}{\line(0,-1){0.25}}
\multiput(91.18,75.84)(0.03,-0.24){1}{\line(0,-1){0.24}}
\multiput(91.15,76.07)(0.03,-0.23){1}{\line(0,-1){0.23}}
\multiput(91.12,76.3)(0.03,-0.23){1}{\line(0,-1){0.23}}
\multiput(91.09,76.52)(0.03,-0.22){1}{\line(0,-1){0.22}}
\multiput(91.05,76.73)(0.03,-0.21){1}{\line(0,-1){0.21}}
\multiput(91.02,76.94)(0.03,-0.21){1}{\line(0,-1){0.21}}
\multiput(90.99,77.14)(0.03,-0.2){1}{\line(0,-1){0.2}}
\multiput(90.96,77.34)(0.03,-0.19){1}{\line(0,-1){0.19}}
\multiput(90.92,77.52)(0.03,-0.19){1}{\line(0,-1){0.19}}
\multiput(90.89,77.71)(0.03,-0.18){1}{\line(0,-1){0.18}}
\multiput(90.86,77.88)(0.03,-0.17){1}{\line(0,-1){0.17}}
\multiput(90.82,78.05)(0.03,-0.17){1}{\line(0,-1){0.17}}
\multiput(90.79,78.21)(0.03,-0.16){1}{\line(0,-1){0.16}}
\multiput(90.76,78.36)(0.03,-0.15){1}{\line(0,-1){0.15}}
\multiput(90.72,78.51)(0.03,-0.15){1}{\line(0,-1){0.15}}
\multiput(90.69,78.65)(0.03,-0.14){1}{\line(0,-1){0.14}}
\multiput(90.65,78.79)(0.03,-0.13){1}{\line(0,-1){0.13}}
\multiput(90.62,78.91)(0.03,-0.13){1}{\line(0,-1){0.13}}
\multiput(90.58,79.03)(0.03,-0.12){1}{\line(0,-1){0.12}}
\multiput(90.55,79.15)(0.03,-0.11){1}{\line(0,-1){0.11}}
\multiput(90.51,79.25)(0.03,-0.11){1}{\line(0,-1){0.11}}
\multiput(90.48,79.35)(0.03,-0.1){1}{\line(0,-1){0.1}}
\multiput(90.44,79.44)(0.04,-0.09){1}{\line(0,-1){0.09}}
\multiput(90.41,79.53)(0.04,-0.09){1}{\line(0,-1){0.09}}
\multiput(90.37,79.61)(0.04,-0.08){1}{\line(0,-1){0.08}}
\multiput(90.34,79.68)(0.04,-0.07){1}{\line(0,-1){0.07}}
\multiput(90.3,79.74)(0.04,-0.06){1}{\line(0,-1){0.06}}
\multiput(90.27,79.8)(0.04,-0.06){1}{\line(0,-1){0.06}}
\multiput(90.23,79.85)(0.04,-0.05){1}{\line(0,-1){0.05}}
\multiput(90.2,79.89)(0.04,-0.04){1}{\line(0,-1){0.04}}
\multiput(90.16,79.93)(0.04,-0.04){1}{\line(0,-1){0.04}}
\multiput(90.12,79.96)(0.04,-0.03){1}{\line(1,0){0.04}}
\multiput(90.09,79.98)(0.04,-0.02){1}{\line(1,0){0.04}}
\multiput(90.05,79.99)(0.04,-0.01){1}{\line(1,0){0.04}}
\multiput(90.02,80)(0.04,-0.01){1}{\line(1,0){0.04}}
\put(89.98,80){\line(1,0){0.04}}
\multiput(89.95,79.99)(0.04,0.01){1}{\line(1,0){0.04}}
\multiput(89.91,79.98)(0.04,0.01){1}{\line(1,0){0.04}}
\multiput(89.88,79.96)(0.04,0.02){1}{\line(1,0){0.04}}
\multiput(89.84,79.93)(0.04,0.03){1}{\line(1,0){0.04}}
\multiput(89.8,79.89)(0.04,0.04){1}{\line(0,1){0.04}}
\multiput(89.77,79.85)(0.04,0.04){1}{\line(0,1){0.04}}
\multiput(89.73,79.8)(0.04,0.05){1}{\line(0,1){0.05}}
\multiput(89.7,79.74)(0.04,0.06){1}{\line(0,1){0.06}}
\multiput(89.66,79.68)(0.04,0.06){1}{\line(0,1){0.06}}
\multiput(89.63,79.61)(0.04,0.07){1}{\line(0,1){0.07}}
\multiput(89.59,79.53)(0.04,0.08){1}{\line(0,1){0.08}}
\multiput(89.56,79.44)(0.04,0.09){1}{\line(0,1){0.09}}
\multiput(89.52,79.35)(0.04,0.09){1}{\line(0,1){0.09}}
\multiput(89.49,79.25)(0.03,0.1){1}{\line(0,1){0.1}}
\multiput(89.45,79.15)(0.03,0.11){1}{\line(0,1){0.11}}
\multiput(89.42,79.03)(0.03,0.11){1}{\line(0,1){0.11}}
\multiput(89.38,78.91)(0.03,0.12){1}{\line(0,1){0.12}}
\multiput(89.35,78.79)(0.03,0.13){1}{\line(0,1){0.13}}
\multiput(89.31,78.65)(0.03,0.13){1}{\line(0,1){0.13}}
\multiput(89.28,78.51)(0.03,0.14){1}{\line(0,1){0.14}}
\multiput(89.24,78.36)(0.03,0.15){1}{\line(0,1){0.15}}
\multiput(89.21,78.21)(0.03,0.15){1}{\line(0,1){0.15}}
\multiput(89.18,78.05)(0.03,0.16){1}{\line(0,1){0.16}}
\multiput(89.14,77.88)(0.03,0.17){1}{\line(0,1){0.17}}
\multiput(89.11,77.71)(0.03,0.17){1}{\line(0,1){0.17}}
\multiput(89.08,77.52)(0.03,0.18){1}{\line(0,1){0.18}}
\multiput(89.04,77.34)(0.03,0.19){1}{\line(0,1){0.19}}
\multiput(89.01,77.14)(0.03,0.19){1}{\line(0,1){0.19}}
\multiput(88.98,76.94)(0.03,0.2){1}{\line(0,1){0.2}}
\multiput(88.95,76.73)(0.03,0.21){1}{\line(0,1){0.21}}
\multiput(88.91,76.52)(0.03,0.21){1}{\line(0,1){0.21}}
\multiput(88.88,76.3)(0.03,0.22){1}{\line(0,1){0.22}}
\multiput(88.85,76.07)(0.03,0.23){1}{\line(0,1){0.23}}
\multiput(88.82,75.84)(0.03,0.23){1}{\line(0,1){0.23}}
\multiput(88.79,75.6)(0.03,0.24){1}{\line(0,1){0.24}}
\multiput(88.76,75.35)(0.03,0.25){1}{\line(0,1){0.25}}
\multiput(88.72,75.1)(0.03,0.25){1}{\line(0,1){0.25}}
\multiput(88.69,74.84)(0.03,0.26){1}{\line(0,1){0.26}}
\multiput(88.66,74.58)(0.03,0.26){1}{\line(0,1){0.26}}
\multiput(88.63,74.31)(0.03,0.27){1}{\line(0,1){0.27}}
\multiput(88.6,74.03)(0.03,0.28){1}{\line(0,1){0.28}}
\multiput(88.57,73.75)(0.03,0.28){1}{\line(0,1){0.28}}
\multiput(88.55,73.46)(0.03,0.29){1}{\line(0,1){0.29}}
\multiput(88.52,73.17)(0.03,0.29){1}{\line(0,1){0.29}}
\multiput(88.49,72.87)(0.03,0.3){1}{\line(0,1){0.3}}
\multiput(88.46,72.56)(0.03,0.31){1}{\line(0,1){0.31}}
\multiput(88.43,72.25)(0.03,0.31){1}{\line(0,1){0.31}}
\multiput(88.4,71.94)(0.03,0.32){1}{\line(0,1){0.32}}
\multiput(88.38,71.61)(0.03,0.32){1}{\line(0,1){0.32}}
\multiput(88.35,71.29)(0.03,0.33){1}{\line(0,1){0.33}}
\multiput(88.32,70.95)(0.03,0.33){1}{\line(0,1){0.33}}
\multiput(88.3,70.62)(0.03,0.34){1}{\line(0,1){0.34}}
\multiput(88.27,70.27)(0.03,0.34){1}{\line(0,1){0.34}}
\multiput(88.24,69.92)(0.03,0.35){1}{\line(0,1){0.35}}
\multiput(88.22,69.57)(0.03,0.35){1}{\line(0,1){0.35}}
\multiput(88.19,69.21)(0.02,0.36){1}{\line(0,1){0.36}}
\multiput(88.17,68.85)(0.02,0.36){1}{\line(0,1){0.36}}
\multiput(88.15,68.48)(0.02,0.37){1}{\line(0,1){0.37}}
\multiput(88.12,68.11)(0.02,0.37){1}{\line(0,1){0.37}}
\multiput(88.1,67.73)(0.02,0.38){1}{\line(0,1){0.38}}
\multiput(88.08,67.35)(0.02,0.38){1}{\line(0,1){0.38}}
\multiput(88.05,66.96)(0.02,0.39){1}{\line(0,1){0.39}}
\multiput(88.03,66.57)(0.02,0.39){1}{\line(0,1){0.39}}
\multiput(88.01,66.17)(0.02,0.4){1}{\line(0,1){0.4}}
\multiput(87.99,65.77)(0.02,0.4){1}{\line(0,1){0.4}}
\multiput(87.97,65.37)(0.02,0.4){1}{\line(0,1){0.4}}
\multiput(87.95,64.96)(0.02,0.41){1}{\line(0,1){0.41}}
\multiput(87.93,64.55)(0.02,0.41){1}{\line(0,1){0.41}}
\multiput(87.91,64.13)(0.02,0.42){1}{\line(0,1){0.42}}
\multiput(87.89,63.71)(0.02,0.42){1}{\line(0,1){0.42}}
\multiput(87.87,63.29)(0.02,0.42){1}{\line(0,1){0.42}}
\multiput(87.85,62.86)(0.02,0.43){1}{\line(0,1){0.43}}
\multiput(87.83,62.43)(0.02,0.43){1}{\line(0,1){0.43}}
\multiput(87.81,61.99)(0.02,0.44){1}{\line(0,1){0.44}}
\multiput(87.8,61.55)(0.02,0.44){1}{\line(0,1){0.44}}
\multiput(87.78,61.11)(0.02,0.44){1}{\line(0,1){0.44}}
\multiput(87.76,60.67)(0.02,0.45){1}{\line(0,1){0.45}}
\multiput(87.75,60.22)(0.02,0.45){1}{\line(0,1){0.45}}
\multiput(87.73,59.77)(0.02,0.45){1}{\line(0,1){0.45}}
\multiput(87.72,59.31)(0.01,0.45){1}{\line(0,1){0.45}}
\multiput(87.7,58.85)(0.01,0.46){1}{\line(0,1){0.46}}
\multiput(87.69,58.39)(0.01,0.46){1}{\line(0,1){0.46}}
\multiput(87.68,57.93)(0.01,0.46){1}{\line(0,1){0.46}}
\multiput(87.66,57.47)(0.01,0.47){1}{\line(0,1){0.47}}
\multiput(87.65,57)(0.01,0.47){1}{\line(0,1){0.47}}
\multiput(87.64,56.53)(0.01,0.47){1}{\line(0,1){0.47}}
\multiput(87.63,56.05)(0.01,0.47){1}{\line(0,1){0.47}}
\multiput(87.62,55.58)(0.01,0.48){1}{\line(0,1){0.48}}
\multiput(87.61,55.1)(0.01,0.48){1}{\line(0,1){0.48}}
\multiput(87.6,54.62)(0.01,0.48){1}{\line(0,1){0.48}}
\multiput(87.59,54.14)(0.01,0.48){1}{\line(0,1){0.48}}
\multiput(87.58,53.66)(0.01,0.48){1}{\line(0,1){0.48}}
\multiput(87.57,53.17)(0.01,0.49){1}{\line(0,1){0.49}}
\multiput(87.56,52.68)(0.01,0.49){1}{\line(0,1){0.49}}
\multiput(87.55,52.2)(0.01,0.49){1}{\line(0,1){0.49}}
\multiput(87.55,51.71)(0.01,0.49){1}{\line(0,1){0.49}}
\multiput(87.54,51.21)(0.01,0.49){1}{\line(0,1){0.49}}
\multiput(87.53,50.72)(0.01,0.49){1}{\line(0,1){0.49}}
\multiput(87.53,50.23)(0.01,0.49){1}{\line(0,1){0.49}}
\multiput(87.52,49.73)(0.01,0.49){1}{\line(0,1){0.49}}
\multiput(87.52,49.24)(0,0.5){1}{\line(0,1){0.5}}
\multiput(87.51,48.74)(0,0.5){1}{\line(0,1){0.5}}
\multiput(87.51,48.24)(0,0.5){1}{\line(0,1){0.5}}
\multiput(87.51,47.75)(0,0.5){1}{\line(0,1){0.5}}
\multiput(87.51,47.25)(0,0.5){1}{\line(0,1){0.5}}
\multiput(87.5,46.75)(0,0.5){1}{\line(0,1){0.5}}
\multiput(87.5,46.25)(0,0.5){1}{\line(0,1){0.5}}
\multiput(87.5,45.75)(0,0.5){1}{\line(0,1){0.5}}
\multiput(87.5,45.25)(0,0.5){1}{\line(0,1){0.5}}
\put(115,45){\makebox(0,0)[cc]{$ r = r_0$}}
\end{picture}}

\begin{document}

\title{Microscopic and Macroscopic Entropy of Extremal Black Holes
in String Theory}

%\titlerunning{Entropy of Extremal Black Holes} 

\author{Ashoke Sen}

\institute{Ashoke Sen\\
Harish-Chandra Research Institute\\
Chhatnag Road, Jhusi\\
Allahabad 211019, India\\
E-mail:   sen@mri.ernet.in}

\maketitle

\begin{abstract}

This is a short review summarizing  
the current status of the comparison between microscopic and macroscopic
entropy of extremal BPS black holes in string theory.

\end{abstract}

\section{Introduction}

About forty years ago Bekenstein and Hawking gave a universal formula for the black 
hole entropy. This takes the form
\be \label{esbhpre}
S_{BH} = {A_H\over 4 G}
\ee
in $\hbar=c=1$ units. Here $A_H$ is the area of the event horizon and $G$ is the Newton's
gravitational constant. 
This suggests that when the curvature at the horizon is small compared to the Planck scale, 
the number $d_{micro}$ of black hole microstates should be given by
\be
d_{micro} \simeq \exp[S_{BH}]\, .
\ee
Our goal in this talk will be to  discuss how this formula is verified in string theory
by direct counting, including possible
corrections to both sides
of the formula. A more detailed review of these results can be found in
\cite{1008.3801}. Other reviews covering somewhat different aspects of the subject
include \cite{0708.1270,1208.4814,1312.7168}.

Counting black hole microstates directly is difficult.
The strategy one follows in string theory is to work with supersymmetric
(BPS) black holes whose ÔdegeneracyÕ is independent of the coupling constant.
We can then
count microstates in the weak coupling limit when gravity can be ignored, 
and compare the result with $\exp[S_{BH}]$, computed when 
gravity is strong enough so that the system can be described as a black hole.
Beginning with the work of Strominger and Vafa\cite{9601029} this has now been achieved
for a wide class of extremal supersymmetric black holes in string theory.
The result takes the form
\be \label{esbh}
S_{BH}(q) \simeq \ln d_{micro}(q)\, ,
\ee
where
$S_{BH}(q)$ is the entropy of a supersymmetric black hole
carrying charges $q$ and 
$d_{micro}(q)$ is the number of supersymmetric
quantum states carrying the same charge, computed
in the weak coupling limit. 
Since typically these theories carry multiple U(1) gauge fields, the black
hole is characterized by more than one charge and hence $q$ stands for a set of charges
$\{q_1, q_2, ...\}$ -- one for each $U(1)$ gauge group present in the low energy theory.
The $\simeq$ symbol used in \refb{esbh} reflects the fact that
in \cite{9601029}, as well as in many follow-up papers, the comparison between 
$S_{BH}(q)$ and $\ln d_{micro}(q)$ is done in the limit of large charges $q$ 
so that the horizon has low curvature, and the
higher derivative terms in the action and quantum gravity corrections can be ignored.
Also in this limit the computation of $d_{micro}(q)$ simplifies. \cite{9711053,9812082}
represent some early papers which
go beyond
this approximation.

In the last paragraph we said that $d_{micro}$ is the number of quantum states,
but this is not quite correct.
In actual practice, what one computes in the  weak coupling limit is an index, \i.e.\
$d_{micro}(q)$ is taken to be the difference between the 
number of bosonic states of charge $q$ and
number of fermionic states of charge $q$.
In 3+1 dimensions this can be expressed 
as\footnote{In this definition we have to factor out the trace over the fermion zero
modes which produce the supermultiplet structure. See \cite{1008.3801} for a detailed
discussion. \label{fo1}}
\be 
d_{micro}(q) =Tr_q[ (-1)^{2J}]\, ,
\ee
where $J$ is the third component of the
angular momentum, and $Tr_q[(-1)^{2J}]$ denotes the 
sum of the expectation value of $(-1)^{2J}$
over all BPS states carrying charge $q$.
It is the index $d_{micro}$, and not the degeneracy, that is independent of the coupling 
constant.

The questions that we shall try to answer in this talk are
\begin{enumerate}
\item  \label{i1}
Can we justify comparing $\exp[S_{BH}]$ , which measures degeneracy, 
with the index computed at weak coupling?
\item Can we find a prescription for computing the exact black hole entropy 
$S_{macro}$, taking 
into account higher derivative and quantum corrections? \label{i2}
\item Assuming that we find affirmative answers to the previous two questions, does
the result for $S_{macro}$ agree with $\ln d_{micro}$?
\item Can macroscopic analysis tell us about other
statistical properties of the microstates besides degeneracy?
\end{enumerate}
Throughout this talk we shall refer to as 
microscopic an analysis where we study the spectrum of supersymmetric 
states by switching 
off gravitational interaction and directly analyzing the quantum states of a system
of solitons of string theory carrying total charge $q$. On the other hand we refer to an analysis as
macroscopic when we take into account the  effect of gravity  and
consequently describe the system as a black hole. In the latter case we cannot directly
identify the quantum states underlying the system, but nevertheless techniques of quantum gravity
and string theory can be used to derive various statistical properties of these quantum states.

Question 2 listed above has  been partially answered by Wald,
who wrote down 
a general formula for the black hole entropy taking into account higher derivative 
terms in the classical effective action\cite{9307038}. The Wald entropy, which we shall denote
by $S_{wald}$, reduces to the Bekenstein-Hawking entropy $S_{BH}$ for Einstein's
theory coupled to ordinary matter field via 2-derivative coupling.  However
quantum corrections to the entropy are not captured by WaldÕs formula. One of our
goals will be to give the quantum corrected formula for the entropy of an extremal black hole.

The plan for the rest of the talk is as follows. First 
we shall give a brief review of what is known about the index on the microscopic
side. Then we shall describe the
general formalism for computing extremal black hole entropy from macroscopic i.e. quantum gravity
analysis, generalizing Bekenstein-Hawking-Wald formula.
Finally, based on this general formula, we shall justify why it makes sense to compare the index
computed on the microscopic side to the degeneracy computed on the macroscopic
side, derive specific predictions from
the macroscopic analysis and test these predictions against microscopic results.
A more detailed review and more results can
be found in \cite{1008.3801}.

\section{Microscopic results} 

Exact microscopic results for the index are known for a wide class of states in a 
wide class of string theories with 16 or 32 unbroken 
supersymmetries. 
These include type II string
theories compactified on $T^6$ and $T^5$, heterotic string theories on $T^6$ and $T^5$ and
special class of orbifolds of these models preserving 16 supersymmetries\cite{9505054}. 
In each case the result is given in terms of Fourier coefficients of some known functions with modular properties.
For theories with 16 unbroken supersymmetries these functions are Siegel modular 
forms\cite{9607026,0510147,0609109,0802.1556,0803.2692} while for
theories with 32 unbroken supersymmetries these functions are given by appropriate weak
Jacobi forms\cite{9903163,0506151}. This allows us to compute the index $d_{micro}$ 
as an exact integer for a given set of charges carried by the black hole.

In each of these examples we can also systematically calculate the behavior of 
$d_{micro}(q)$ for large charges.
General strategy one follows is to compute the Fourier integrals using saddle point method.
The leading term of $\ln d_{micro}$ always agrees with the Bekenstein-Hawking 
entropy of an extremal black hole with the same charge. This is an old story by 
now\cite{9601029}, and
will not be discussed in detail.
Our goal will be to understand to what extent macroscopic 
analysis can explain the subleading terms in $\ln d_{micro}$.

As an example we can consider type IIB string theory on $T^6$. At a generic point in the
moduli space this theory has several $U(1)$ gauge fields. Let $\{Q_i\}$ and $\{P_i\}$
denote the electric and magnetic charges carried by a supersymmetric black hole under these U(1)
gauge fields. One finds that for $\{Q_i\}$, $\{P_i\}$ satisfying certain restrictions, including
$\gcd(Q^2/2,P^2/2,Q\cdot P)=1$,
the index is given by\cite{0506151}
\be \label{edmc}
 d_{micro} =  (-1)^{Q\cdot P+1}\,
 c(\Delta),  \qquad \Delta  \equiv Q^2 P^2 - (Q\cdot P)^2\, ,
 \ee
 where $Q^2$, $P^2$ and $Q\cdot P$ are certain bilinear combinations of charges, and
 $c(u)$ is defined through
 \be
 -\vartheta_1(z|\tau)^2 \, \eta(\tau)^{-6} \equiv \sum_{k,l}  c(4k-l^2)\, 
e^{2\pi i (k\tau+l z)}\, .
\ee
Here $\vartheta_1(z|\tau)$ is the odd Jacobi theta function and $\eta(\tau)$
is the Dedekind eta function. Since $\vartheta_1(z|\tau)$ and $\eta(\tau)$ have
known Fourier expansion coefficients, \refb{edmc} gives a complete expression for the
index as a function of the charges.

One can compute the index for large charges by evaluating the Fourier integral using saddle
point method. One finds the result\cite{0908.0039}
\be
\ln d_{micro} = \pi \sqrt\Delta { - 2 \, \ln\, } \Delta + \cdots \, .
\ee
On the other hand the Bekenstein-Hawking entropy of the same black hole,
computed from \refb{esbhpre}, is given by
\be
\pi \sqrt\Delta\, .
\ee
Thus we see that the microscopic result for $\ln d_{micro}$ agrees with $S_{BH}$ for large
$\Delta$. 
Later we shall see the origin of the logarithmic term
from the macroscopic side.
The results for other theories are similar but the details are different. 

In many of these theories we can also define twisted index.
Suppose the theory has a discrete
$ \ZZZ_N$ symmetry generated by $g$ which commutes with the unbroken supersymmetries
of the state.
Then we define
\be \label{etwistdef}
d_{micro}^{(g)}\equiv Tr_q\left[(-1)^{2J} g\right],
\ee
where $Tr_q$ as usual denotes sum of the expectation value 
over BPS states of charge $q$.
Thus
$d_{micro}^{(g)}$ contains information about the distribution of  the eigenvalues 
of the $\ZZZ_N$ generator among the
microstates.
It can be argued that $d_{micro}^{(g)}$ 
is independent of the coupling constant. Thus the microscopic and macroscopic results for
these quantities can be compared. On the microscopic side, 
these quantities can be computed in this class of theories with 16 or 32 supercharges
and the results are again given by Fourier coefficients
of appropriate Siegel modular forms\cite{0911.1563,1002.3857}. 
Analysis of the large charge limit of these quantities shows that they
grow as
\be \label{ek1}
\exp\left[S_{BH}/N\right]
\ee
where $S_{BH}$ is the Bekenstein-Hawking entropy of the corresponding black hole.
The fact that $d_{micro}^{(g)}$ is exponentially smaller that $d_{micro}$ indicates that
there is a large degree of cancellation between the contribution to $d_{micro}^{(g)}$ from
different microstates. This is somewhat surprising from the microscopic viewpoint, but we shall
see that this has a simple explanation from the macroscopic side.

All these results (and some more to be discussed in \S\ref{s4})
provide us with a wealth of
`experimental data' against which we can test the
predictions of macroscopic analysis based on quantum gravity or string theory.
In the rest of the talk we shall describe to what extent we have achieved this goal.

\section{Macroscopic analysis}

Our analysis on the macroscopic side begins with the observation that 
supersymmetric black holes in string theory are extremal.
In the extremal limit a  black hole
acquires an infinite throat described by an $ AdS_2$ factor, separating the
horizon from the asymptotic space-time.\footnote{A detailed review of this and its consequences
in the classical theory,
as well as the relevant references can be found in \cite{0708.1270}.}
The full throat geometry takes the form $ AdS_2\times K$ where 
$K$ includes the angular coordinates of space, e.g. polar and
azimuthal angles in 3+1 dimensions,
 as well as the internal space on which we 
 have compactified string theory.
The metric on the throat is given by
\be 
ds^2 =  a^2 \left[-r^2 dt^2 + 
   {dr^2\over r^2} \right]+ ds_K^2
\ee
where $ds_K^2$ denotes the metric on $K$.
In this coordinate system the original horizon is towards  $r\to 0$ whereas the 
original asymptotic space-time is towards $r\to\infty$.

\begin{figure}
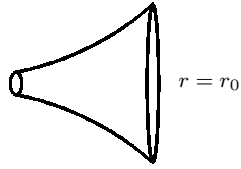
\begin{center}
\vbox{ ~
\vskip -190pt
\figadssmall
}
\end{center}
\caption{Euclidean geometry described by the $(r,\theta)$ coordinates
in  \refb{euclid} with an IR cut-off $r\le r_0$. \label{fa}}
\end{figure}

Our strategy will be to analyze euclidean path integral on this space-time.
For this we take Euclidean continuation of the throat metric, \i.e.\ replace $t$ by $-i\theta$.
This gives
\be \label{euclid}
ds^2 =
a^2\left( r^2  \, d\theta^2+{dr^2\over
r^2}\right) + ds_K^2\, .
\ee
Since we can change the period of $\theta$ by a rescaling
$r\to \lambda r, \theta\to \theta/\lambda$, we use this freedom
to fix the period of $\theta$ to be
$2\pi$. The boundary of Euclidean $AdS_2$ is towards $r\to\infty$.
Following usual practice we
regularize the infinite volume of $
AdS_2$ by putting a cut-off $r\le r_0$ (see fig~\ref{fa}).
This makes the $AdS_2$ boundary have a 
finite length
\be
L=2\pi \, a\, r_0\, .
\ee
Eventually we need to take the limit $r_0\to \infty$, which in turn corresponds to
$L\to \infty$.

We now proceed as follows:
\begin{enumerate}
\item Define the partition function:
\be \label{epart}
Z= \int D\varphi \, \exp[-\hbox{Action}
 - \hbox{boundary terms}]
 \ee
 where $\varphi$ stands for all fields in the string theory under consideration. We impose the
 boundary condition that for large $r$ the field configuration should approach
the throat geometry of the black hole. Subject to this boundary condition we integrate
over all field configurations and topologies.
\item
Now using the standard AdS/CFT dictionary, 
$Z$ can be reinterpreted as the partition function of a dual quantum mechanical 
system sitting at the boundary $r=r_0$ on a Euclidean time circle of length $L$.
It turns out that the due to peculiar boundary conditions forced on us by the geometry
of $AdS_2$, the Hilbert space of
this dual quantum mechanics contains states carrying fixed charges and
angular momenta\cite{0809.3304}. 
If we denote by $H$ the Hamiltonian of this system, by $E_0$  the ground state
energy and by $d_0$ the ground state degeneracy, then we have
\be \label{ecft1}
Z= Tr(e^{-LH}) \to d_{0}\, e^{-L\, E_0} \quad \hbox{as} \quad L\to \infty\, .
\ee
To arrive at the last expression 
we have assumed that the spectrum has a gap separating the ground state
from the first excited state. Since by the rules of AdS/CFT correspondence 
the dual quantum mechanics should be identified with the one describing the low energy
dynamics of the black hole and since the microscopic description of the
corresponding black hole in string theory always has a discrete spectrum, this is
a reasonable assumption.
\item We now identify the quantum corrected 
 entropy as
\be \label{esmacro}
S_{macro} = \ln \, d_0 =\lim_{L\to\infty} \left(1 - L{d\over dL}\right) 
\ln \, Z \, ,
\ee
where we have used \refb{ecft1} in the last step.
This gives the 
quantum generalization of the Bekenstein - Hawking - Wald formula.
\end{enumerate}

For consistency check we need to verify that in the classical limit $S_{macro}$ defined
above reduces to the Wald entropy $S_{wald}$.
For this we evaluate $Z$ using saddle point approximation. It turns out that the
dominant saddle point that contributes to  $Z$
is the euclidean black hole in $AdS_2$ described by the metric
 \ben \label{eclass}
 ds^2 &=&  a^2 \left[(r^2-1) d\theta^2 + 
   {dr^2\over r^2-1} \right]+
 ds_K^2 \nonumber \\
&=& a^2(d\eta^2 + \sinh^2 \eta \, d\theta^2) + ds_K^2, \qquad 
\eta \equiv \cosh^{-1} r \, .
  \een
This is an allowed saddle point since for large $r$ this approaches the metric
given in \refb{euclid}.
The space spanned by $(\eta,\theta)$ clearly has the topology of a disk. 
The leading contribution to $Z$ is given by $\exp[S_{cl}]$ where $S_{cl}$ denotes the
classical action evaluated in this background. It can be shown that  $S_{macro}$
computed from \refb{esmacro} using this classical value of $Z$ is exactly equal to the Wald entropy
irrespective of the form of the action, as long as the action is invariant under general coordinate
transformation\cite{0809.3304}. 

\section{Comparing macroscopic and microscopic entropy}  \label{s4}

So far we have discussed two independent computations -- one from the
microscopic side by counting of states and the other from the macroscopic side by
performing path integral in the underlying quantum theory of gravity.
We shall now compare the macroscopic and microscopic results.

\subsection{ Degeneracy vs.  index}

As already emphasized, in the microscopic theory we compute an index while the macroscopic
computation is expected to yield degeneracy. We shall now discuss why it is sensible to compare
the two. For simplicity we focus on four dimensional black holes, but the analysis can be
generalized to five dimensional black holes as well.

We begin with the observation that $AdS_2$ space has SL(2,R) isometry.
The supersymmetric black holes we consider are also invariant under 4 supersymmetries.
It turms out that the closure of the symmetry algebra requires many more generators,
and leads to what is  known as $su(1,1|2)$ algebra. The corresponding symmetry group 
contains an $SU(2)$ subgroup which can be identified
as the spatial rotation group. 
Thus supersymmetric black holes must be spherically symmetric.
This in turn implies that they must carry vanishing average angular momentum $\vec J$.

We now combine this with another result that follows from the geometry of $AdS_2$.
As mentioned above \refb{ecft1}, 
the allowed boundary conditions on the gauge fields in $AdS_2$ are such
that $Z$ defined in \refb{epart} is the partition function of
an ensemble of fixed charges and angular momenta 
rather than fixed chemical potentials\cite{0809.3304}, 
and hence  an
extremal black hole describes a microcanonical ensemble of states with all
states carrying same $\vec J$ and other charges\cite{0903.1477,1009.3226}.
Since the average angular momentum is zero, we see that 
all the states in the ensemble described by the black hole must carry zero angular
momentum. Put another way, when the coupling is such that the black hole description is suitable,
all BPS states carry zero angular momentum.\footnote{This argument requires us to 
factor out the contribution of the
goldstino fermion zero modes associated with the supersymmetry generators that are
broken by the black hole. These live outside the event horizon of the black hole and are
not included in the partition function $Z$ computed using $AdS_2$ boundary 
condition\cite{0901.0359,0907.0593}.
This is related to the subtlety in the definition of the index mentioned in footnote \ref{fo1}.}
This gives, for extremal black holes
\be \label{eindex}
\hbox{Index} = Tr_{q} [(-1)^{2J}] =Tr_{q} [1] = \hbox{degeneracy}
= \exp[S_{macro}]\, .
\ee
Thus the exponential of the macroscopic entropy of the supersymmetric black hole 
also computes the index.
Since the index is protected from quantum corrections, it is sensible to compare this
with the
microscopic index computed in a different regime of the coupling constant space.

Besides explaining why comparison of microscopic index with macroscopic degeneracy is 
sensible, this also  has a non-trivial consequence. Since on the macroscopic side the index has
been argued to be equal to degeneracy, the index must be positive. Thus the equality of microscopic
and macroscopic index implies that the microscopic index must also be positive. Now on the
microscopic side the spectrum contains both bosonic and fermionic states, and in general there
is no independent reason as to why the index should be positive, \i.e.\ why
the number of bosonic states should
always be greater than the number of fermionic states. Thus testing the positivity of the microscopic
index is a non-trivial test of the equality of the microscopic and the macroscopic calculations.

\begin{table}
\begin{center}
\caption{Some sample results 
for the microscopic index of single centered black holes in heterotic string theory
on $T^6$. \label{t1}}
\def\st{\vrule height 3ex width 0ex}
\begin{tabular}{|l|l|l|l|l|l|l|l|l|l|l|} \hline 
$(Q^2,P^2){\backslash} Q\cdot P$  & 2 & 3 & 4 & 5 & 6 & 7
\st\\[1ex] \hline \hline
(2,2)   
&  648 & { 0} & { 0} & { 0} & { 0} 
& { 0} 
\st\\[1ex] \hline
(2,4) 
&  { 50064} & { 0} & { 0} & { 0}
& { 0} & { 0}
\st\\[1ex] \hline
(2,6)   &{ 1127472} & 25353  & { 0} 
& { 0} & { 0} & { 0} \st\\[1ex] \hline
(4,4)  &  { 3859456}
&  { 561576} & 12800 & { 0} & { 0} 
& { 0} \st\\[1ex] \hline
(4,6)   & 
{ 110910300}  &  
{ 18458000} 
&  { 1127472} & { 0} & { 0} & { 0}
\st\\[1ex] \hline
(6,6) & 
{ 4173501828}
&  { 920577636} & { 110910300} & { 8533821} & 
{ 153900}  & { 0}
\st\\[1ex] \hline
(2,10) & 185738352 & 16844421 & 
16491600 & {0} & {0} & 
{0} \st\\[1ex]\hline
 \hline 
\end{tabular} 

\end{center} 

\end{table}

So far in all cases where the microscopic results are known, this has been 
verified\cite{1008.4209}. We show in 
table \ref{t1} the microscopic index of a class of black holes in heterotic string theory compactified
on a six dimensional torus $T^6$.\footnote{These coefficients are 
computed from the Fourier expansion coefficients of the inverse
of the Igusa cusp form\cite{igusa}.}
We can see that all the entries are positive, in agreement with the
macroscopic prediction. For arriving at this table we have to carefully remove the contribution
to the index from multi-centered black holes since the latter do not enjoy positivity property.
The positivity of the microscopic index in this theory 
has now been proved  for all values of $Q^2$ 
and $Q\cdot P$ with $P^2=2$ and $P^2=4$\cite{1208.3476} but the full proof is still awaited.

The zero entries in table \ref{t1} occur for $Q^2P^2 < (Q\cdot P)^2$ where there
are no single centered black hole solutions and hence $\exp[S_{macro}]$
vanishes\cite{0702150,1104.1498}. 
The vanishing of the microscopic result for
$Q^2P^2 < (Q\cdot P)^2$
has now been proved for all values of
$Q^2,P^2$ and $Q\cdot P$\cite{1104.1498,1210.4385}, not only in heterotic string theory
on $T^6$ but also in many other models with 16 supersymmetries. This provides another 
non-trivial test of the equality of
macroscopic and microscopic results.

\subsection{Logarithmic 
corrections to  entropy}

Typically the leading entropy $ S_{BH}$ is a homogeneous function
of the various charges $ q_i$. For example in $3+1$ dimensions 
\be
S_{BH}(\Lambda q) = \Lambda^2 \, S_{BH}(q)
\ee
for any $\Lambda$.
Logarithmic corrections refer to subleading correction to the entropy $\propto \ln \Lambda$
in the limit of large $\Lambda$.
These arise from one loop correction to the
leading saddle point result for $ Z_{}$ from loops of
{\it massless} fields\cite{1005.3044,1106.0080}.

\begin{table}
\caption{Macroscopic results for the logarithmic
correction to the extremal black hole entropy. The first column gives the theory in which we 
do the analysis with $\NN$ denoting the number of supersymmetries and $D$ denoting the 
number of space-time
dimensions, 
the second column gives the coefficient of the $\ln \Lambda$ term when the 
various charges are scaled by $\Lambda$, and the last
column gives the result of comparison with the microscopic results. A $\surd$ indicates
matching with the microscopic results, a x denotes disagreement between microscopic and
macroscopic results and a $?$ denotes that the microscopic results are not yet known. Absence
of any x in the last column shows perfect agreement between microscopic and macroscopic results
whenever the results are known on both sides.
\label{t2}}
\begin{center}\def\st{\vrule height 3ex width 0ex}
\begin{tabular}{|l|l|l|l|l|l|l|l|l|l|l|} \hline
The theory  &  logarithmic contribution & microscopic
\st\\[1ex] \hline \hline
$\NN=4$ in D=4 with $ n_v$ matter   &  0 & $\surd$
\st\\[1ex] \hline
$\NN=8$ in D=4 
& $-{ 8\, \ln}\, \Lambda$ & $\surd$
\st\\[1ex] 
\hline
$\NN=2$ in D=4 with $ n_V$ vector 
&  ${ {1\over 6} (23 + n_H - n_V)\, \ln}\,  \Lambda$
& ?
\st\\%[1ex] \hline
and $ n_H$ hyper &   
&  
\st\\[1ex] \hline
$\NN=6$ in D=4 
 & ${ -4\, \ln} \, \Lambda$ & ?
\st\\[1ex] \hline
$\NN=5$ in D=4 
 & ${ -2\, \ln} \, \Lambda$ & ?
\st\\[1ex] \hline
$\NN=3$ with $ n_v$ matter in D=4 & 
 ${ 2\, \ln} \, \Lambda$ & ?
\st\\[1ex] \hline\hline
BMPV in type IIB on $ T^5/\ZZZ_N$ & 
 ${ -{1\over 4} (n_V-3) \, \ln} \, \Lambda$ & $\surd$
\st\\[1ex] %\hline
or $ K3\times S^1/\ZZZ_N$ with $ n_V$ vectors & 
 & 
\st\\[1ex] 
and $\vec J\sim  \Lambda^{3/2}$ & 
 & 
\st\\[1ex] \hline
BMPV in type IIB on $ T^5/\ZZZ_N$  & 
 ${ -{1\over 4} (n_V+3) \, \ln} \, \Lambda$ & $\surd$
\st\\[1ex] %\hline
or $ K3\times S^1/\ZZZ_N$ with $ n_V$ vectors & & 
\st\\[1ex] 
and $\vec J=0$ & & 
\st\\[1ex]\hline
\end{tabular}
\end{center}

\end{table}

Consider a spherically symmetric 
extremal black hole in $D=4$ with horizon size $a$.
The leading contribution to $Z$ defined in \refb{epart} comes from the saddle point 
associated with the euclidean black hole near horizon
geometry
\be
ds^2 = a^2\left( {dr^2\over r^2-1} + (r^2-1) d\theta^2 + 
d\psi^2 + \sin^2\psi d\phi^2\right) + ds_{\rm{compact}}^2 \, ,
\ee
where $ds_{\rm{compact}}^2$ is the metric of the compact space.
Note that unlike in \refb{euclid} where $ds_K^2$ also included the
angular coordinates $(\psi,\phi)$, we have now separated the angular coordinates from
$ds_{\rm compact}^2$ in order to display that
the size of the angular directions is of the same order $a$ as that of $AdS_2$.
When the charges scale uniformly  by a large number $\Lambda$
then $a$ scales as $\Lambda$ 
keeping other parameters, e.g.
$ds_{\rm {compact}}^2$ and vacuum expectation values of various scalar fields, 
fixed.
If $\Delta$ denotes the kinetic operator  of the four dimensional massless fields
in the near horizon background then the one loop contribution to $ Z_{}$ from
the non-zero modes is given by
\be
({\rm sdet}' \Delta )^{-1/2} = \exp\bigg[-{1\over 2} \ln \, {\rm sdet}'\Delta\bigg]
\ee
where ${\rm sdet}'$ denotes the super-determinant sans the contribution from the zero
modes of $\Delta$. 
We can use heat kernel expansion to determine the terms in ${\ln \, {\rm sdet}'}\, \Delta$
proportional to $\ln  a$. These arise from modes with eigenvalues $<<  m_{pl}^2$ and hence are
insensitive to the ultraviolet cut-off.

To compute the full logarithmic correction to $S_{macro}$ we must also compute the contribution
from the zero modes of $\Delta$. Since there is no damping factor from
the $\exp[-\hbox{Action}]$ term in the integrand of \refb{epart} at the quadratic level,
integration over the zero modes gives the volume of the space spanned by these modes
rather than a determinant.
By careful analysis we can determine the dependence of this volume on the 
size $a$ of $ AdS_2$ and $ S^2$.
The final result is obtained by combining the zero mode and the non-zero mode
contributions.

Table \ref{t2} shows the final result for the logarithmic correction to the macroscopic entropy
for a wide variety of theories\cite{1005.3044,1106.0080,1108.3842,1109.0444}. 
As can be seen from the explanation provided in the caption,
in all cases where the microscopic results are known there is perfect agreement between 
microscopic and macroscopic results. 

\subsection{Twisted index}

We can also compute the twisted index defined in \refb{etwistdef} from the macroscopic
side.
By our earlier argument on the macroscopic side we have
\be 
Tr_q[ (-1)^{2J} g]= Tr_q[ g ]
\ee
since supersymmetric black holes carry zero angular momentum. 
We can compute this in the same way as the degeneracy by
evaluating the partition function 
\be \label{etwistpart}
Z_g= \int [D\varphi]_g \, \exp[-\hbox{Action}
 - \hbox{boundary terms}]
 \ee
where $[D\varphi]_g$ means that we perform path integral over all the fields 
$\varphi$ by imposing a $g$-twisted boundary condition
along the boundary circle of $AdS_2$. 
In the dual quantum mechanics this corresponds to $Tr[\exp(-L\, H) g]$.
In the $L\to\infty$ limit only the contribution from the ground state remains, and
the result is given by 
\be
\exp[-L\, E_0]\, d_{macro}^{(g)}\, ,
\ee
where $d_{macro}^{(g)}$ is the trace of $g$ over ground states. From this
we can derive an expression for $d_{macro}^{(g)}$ analogous to
the one for $S_{macro}$ in  \refb{esmacro}:
\be \label{etindex}
\ln \, d_{macro}^{(g)} = \lim_{L\to\infty} \left( 1 - L {d\over d L}\right) \ln \, Z_g
\, .
\ee
We can regard $d_{macro}^{(g)}$ given above as the macroscopic result for the twisted index.

For evaluating the leading contribution to 
$Z_g$ we note that the original Euclidean black hole solution given in
\refb{eclass} is no longer an allowed saddle point of the path integral
\refb{etwistpart}, since the boundary circle at $\eta=\eta_0$
can be contracted by deforming it to $\eta=0$ and hence a $g$-twisted boundary condition will
lead to singular boundary condition at $\eta=0$. 
However there is a different saddle point, related to the original saddle
point by a $\ZZZ_N$ orbifolding,  that gives the dominant contribution 
to the path integral\cite{0911.1563}.
The metric associated with this saddle point is given by\footnote{By replacing $r$ by $r/N$ and
$\theta$ by $N\theta$ we can bring the metric \refb{etwistfin} to 
the form given in \refb{eclass}. However since the redefined $\theta$ has period $2\pi/N$, the
metric 
has a $\ZZZ_N$ orbifold singularity at the origin $\eta=0$. 
After including the effect of the $g$-twist, and a $\ZZZ_N$ rotation on $S^2$ that accompanies
the $\theta$ translation by $2\pi/N$, one can show that the orbifold singularity at the
origin is an allowed supersymmetric defect in string theory\cite{0911.1563}.}
\be \label{etwistfin}
ds^2 =  a^2 \left[(r^2-N^{-2}) d\theta^2 + 
   {dr^2\over r^2-N^{-2}} \right]+
 ds_K^2\, .
 \ee
The corresponding classical action can be used to evaluate the saddle point contribution
to \refb{etindex}. The leading order result for $d_{macro}^{(g)}$ computed from
\refb{eindex} is given by\cite{0911.1563}
\be \label{emacrotwist}
d_{macro}^{(g)} = \exp[S_{wald}/N] \, ,
\ee
where $S_{wald}$ denotes the Wald entropy of the same black hole. 
The factor of $1/N$ multiplying $S_{wald}$
can be easily understood as the result of $\ZZZ_N$ orbifolding.
Eq.\refb{emacrotwist}
is in perfect agreement with \refb{ek1}.

This provides us with a non-trivial test of the equality between
the microscopic and the macroscopic results for the twisted index.  
It should in principle be possible to compute
the logarithmic correction to the twisted index and compare the microscopic and macroscopic results.
Recently a close cousin of this computation has been performed\cite{1311.6286} where the authors
study logarithmic correction to the exponentially suppressed corrections to the untwisted index.
On the macroscopic side these corrections come from subleading saddle points of the path integral
given by the $\ZZZ_N$ orbifold of the euclidean black hole metric \refb{euclid}, but without
any $g$-twist\cite{0810.3472}.

\subsection{Non-logarithmic one loop corrections}

As can be seen from table \ref{t2}, the $\NN=4$ supersymmetric string theories in four
space-time dimensions do not have any logarithmic corrections to their entropy. 
However these theories do have finite corrections of order unity to the entropy. 
On the microscopic side these take the
form
\be \label{esk1}
\ln d_{micro} =  \pi \sqrt{Q^2 P^2 - (Q\cdot P)^2} +
f\left( {Q\cdot P\over P^2}, {\sqrt{Q^2 P^2 - (Q\cdot P)^2}
\over P^2}\right)+ \OO( \hbox{charge}^{-2})\, 
\ee
where $f$ is a known function involving modular forms. 
Of these the universal first term
$\pi  \sqrt{Q^2 P^2 - (Q\cdot P)^2} $ represents the Bekenstein-Hawking entropy of the same
black hole. Thus the question is: is it possible to reproduce the second term from the
macroscopic side? Note that this term is non-universal since the function $f$ differs from
one theory to another.

It turns out that on the macroscopic side, one loop correction produces
a special class of terms in the effective action, given by
\be \label{esk2}
-{1\over 64\pi^2}  \int d^4 x \sqrt{g} \, f(a,S)\, 
\left\{ R_{\mu\nu\rho\sigma} R^{\mu\nu\rho\sigma}
- 4 R_{\mu\nu} R^{\mu\nu}
+ R^2
\right\}
\ee
where $(a,S)$ denote the axion-dilaton field, $R_{\mu\nu\rho\sigma}$ is the Riemann tensor
and $f$ is the same function that appeared in \refb{esk1}. 
One can compute the effect of this term on black hole entropy, and finds that it precisely
reproduces the result given in \refb{esk1}. This was first checked in heterotic string
theory on $T^6$ in \cite{0412287} and later in various other $\NN=4$ supersymmetric
string theories in \cite{0510147,0609109}. 
These tests are quite non-trivial since on the macroscopic
side computation of the effective action requires carefully evaluating the contribution from
the massive string states propagating in the loop.  Furthermore for different string
compactifications we get different functions $f$, and in each case the function $f$ appearing
in the effective action coincides with the one that arises from counting of black hole
microstates.

We should however add a word of caution  here. Eq.\refb{esk2} gives only one set of terms in the
effective action which are generated to this order. There are also other terms which could
contribute to the entropy and destroy the equality between the macroscopic and the microscopic
results. It is generally expected that there are appropriate non-renormalization theorems which
prevent these other terms from contributing to the entropy, but no such theorem has been proved
to this date. The current status of this study as well as further references can be found in
\cite{1401.6591}.

\section{Conclusion}

In this talk we have described a systematic procedure for computing the macroscopic 
entropy of an extremal black hole taking into account both higher derivative and quantum
corrections. We also reviewed the current status of the comparison between microscopic and 
macroscopic results for the extremal black hole entropy in string theory. 
This comparison includes
\begin{itemize}
\item sign of the index,
\item logarithmic correction to the entropy,
\item asymptotic growth of the twisted index, and
\item finite corrections to the entropy.
\end{itemize}
We have seen that so
far string theory has passed all tests. This gives us confidence in the internal consistency of
string theory and shows that string theory is indeed a strong candidate for a quantum theory
of gravity.

We have not been able to review all aspects of the subject. In particular there have been considerable
amount of progress in recent years in evaluating the path integral \refb{epart} using
localization techniques\cite{0905.2686,1012.0265,1111.1161,1208.6221}. 
So far the results are encouraging but more work needs to be done.
If successful, this could lead us to an exact expression for the index from the macroscopic side
which can then be compared with the corresponding exact results from the microscopic side.

Finally we must mention that the methods discussed here are also applicable for computing 
logarithmic corrections to the entropy
of non-supersymmetric extremal black holes\cite{1109.3706,1204.4061}. 
(Ref.\cite{9709064} represents early work on this
subject). For example entropy of extremal Kerr black hole in pure gravity has a correction
\be
{16\over 45} \ln A_H\, ,
\ee
where $A_H$ is the area of the event horizon.
It is a challenge for any theory of quantum gravity (in particular Kerr/CFT 
correspondence\cite{0809.4266} which
attempts to give a CFT dual description of extremal Kerr black holes) to reproduce this result.
A generalization of these techniques can also be used to compute logarithmic corrections to the
macroscopic entropy of a Schwarzschild black hole. The result is\cite{1205.0971}
(see \cite{9407001,9408068,9412161} for earlier work along this line)
\be
{77\over 90} \, \ln \, A_H\, .
\ee
Again it is a challenge for any theory of quantum gravity to reproduce this result. The
complexity of the coefficients indicate that there may not be a simple microscopic
derivation of these
results.

\bigskip

\noindent {\bf Acknowledgement:}
This work was
supported in part by the 
DAE project 12-R\&D-HRI-5.02-0303 and the
J.~C.~Bose fellowship of 
the Department of Science and Technology, India.

\end{document}